\documentclass[sn-mathphys,Numbered]{sn-jnl}

\usepackage{graphicx}%
\usepackage{multirow}%
\usepackage{amsmath,amssymb,amsfonts}%
\usepackage{amsthm}%
\usepackage{mathrsfs}%
\usepackage[title]{appendix}%
\usepackage{xcolor}%
\usepackage{textcomp}%
\usepackage{manyfoot}%
\usepackage{booktabs}%
\usepackage{algorithm}%
\usepackage{algorithmicx}%
\usepackage{algpseudocode}%
\usepackage{listings}%
\usepackage[numbers]{natbib}
\usepackage{subcaption}
\usepackage{epsfig,amsbsy}
\numberwithin{equation}{section}

\newtheorem{remark}{Remark}%

\newcommand{\err}{\text{\normalfont err}}

\begin{document}

\title[Multi-grid Stochastic reaction--diffusion Modelling]{Multi-grid reaction--diffusion master equation: applications to morphogen gradient modelling}

\author[1]{\fnm{Radek} \sur{Erban}}\email{erban@maths.ox.ac.uk}

\author[2]{\fnm{Stefanie} \sur{Winkelmann}}\email{winkelmann@zib.de}

\affil[1]{\orgdiv{Mathematical Institute}, \orgname{University of Oxford}, \orgaddress{\street{Radcliffe Observatory Quarter
Woodstock Road}, \city{Oxford}, \postcode{OX2 6GG}, \country{United Kingdom}}}

\affil[2]{\orgname{Zuse Institute Berlin (ZIB)}, \orgaddress{\street{Takustrasse 7}, \city{Berlin}, \postcode{D-14195}, \country{Germany}}}

\abstract{The multi-grid reaction-diffusion master equation (mgRDME) provides a generalization of stochastic compartment-based reaction-diffusion modelling described by the standard reaction-diffusion master equation (RDME). By enabling different resolutions on lattices for biochemical species with different diffusion constants, the mgRDME approach improves both accuracy and efficiency of compartment-based reaction-diffusion simulations. The mgRDME framework is examined through its application to morphogen gradient formation in stochastic reaction-diffusion scenarios, using both an analytically tractable first-order reaction network and a model with a second-order reaction. The results obtained by the mgRDME modelling are compared with the standard RDME model and with the (more detailed) particle-based Brownian dynamics simulations. The dependence of error and numerical cost on the compartment sizes is defined and investigated through a multi-objective optimization problem.}


\maketitle

\section{Introduction}\label{sec1}

Compartment-based stochastic reaction-diffusion models have been used for modelling a range of biological processes, including Min-protein oscillations~\cite{Fange:2006:NMP} and  ribosome biogenesis~\cite{Earnest:2015:TWM} in {\it E. coli}, 
calcium signalling~\cite{Denizot:2019:SCS,Dobramysl:2016:PMM}, gene expression~\cite{winkelmann2016spatiotemporal}, actin transport in filopodia~\cite{Zhuravlev:2010:DAT,Erban:2014:MSR} and epidemic spreading~\cite{winkelmann2021mathematical}. They can be simulated using algorithms for continuous-time discrete-space Markov chains and are mathematically described using the reaction-diffusion master equation (RDME) for the probability mass function~\cite{erban2020stochastic,winkelmann2020stochastic}. At the analytical level, the RDME builds a bridge between microscopic models for reaction-diffusion processes and macroscopic approximations through partial differential equations~\cite{Flegg:2012:TRM,montefusco2023route,Flegg:2015:CMC,Kang:2019:MSR}.

To formulate a standard compartment-based model, the computational domain is discretized into compartments and diffusion is modelled as a jump process between the compartments~\cite{winkelmann2020stochastic}. Considering three-dimensional domains, it has been shown that the compartment size cannot be chosen arbitrarily small for systems containing second-order or higher-order reactions, {\it i.e.}, the error of a compartment-based simulation increases as the compartment size approaches zero~\cite{erban2020stochastic}. There have been numerous studies~\cite{isaacson2009reaction,erban2009stochastic,hellander2012reaction,kang2012new,isaacson2013convergent,agbanusi2014comparison,hellander2016reaction,isaacson2018unstructured} on the optimal choice of the compartment size and its influence on the approximation quality compared to microscopic modeling approaches given by the Doi model~\cite{doi1976stochastic,doi1976second} or the Smoluchowski model~\cite{andrews2010detailed,andrews2012spatial}.

The optimal size of the compartment depends on the diffusion constant~\cite{erban2009stochastic,isaacson2009reaction,hellander2012reaction,kang2012new,isaacson2013convergent,agbanusi2014comparison,hellander2016reaction,isaacson2018unstructured}.
In particular, if a biological system consists of molecular species with different diffusion constants, the compartment-based model can be naturally generalized to allow for different meshes (compartment sizes) for different chemical species~\cite{cao2014stochastic,Hellander:2020:HAR,li2012multiscale}. In this paper, we analyze such models using a generalization of the RDME which we will call the \textit{multi-grid reaction diffusion master equation} (mgRDME). Numerical simulations of multi-grid reaction-diffusion models allow for high accuracy at reduced numerical cost compared to fully microscopically resolved systems~
\cite{cao2014stochastic,Chatterjee:2004:SAL,Dai:2008:CGL}. In some computational frameworks~\cite{Hellander:2020:HAR}, molecules can transfer from a fine-grained mesh to a coarse-grained mesh whenever appropriate. In the mgRDME that will be analyzed in this work, the grid size remains constant for each species but varies across species. 

The mgRDME framework will be analyzed by applying it to biochemical systems of morphogen gradient formation~\cite{saunders2009pays,kicheva2007kinetics,Bergmann:2007:PDB,Gregor:2007:SND}. Morphogens are signaling molecules whose non-uniform distribution control pattern formation (or morphogenesis) during the development of multicellular organisms. Starting with the pioneering work of Turing~\cite{Turing:1952:CBM}, it has been shown that pattern formation can naturally arise in reaction-diffusion systems through diffusion-driven instability, provided that the system contains at least two chemical species with different diffusion constants~\cite{Murray:2002:MB2}. In particular, stochastic modelling of Turing patterns is a natural application area of the mgRDME framework~\cite{cao2014stochastic}. 

In this paper, we will focus on pattern formation (morphogen gradient formation) which results from pre-patterning, {\it i.e.}, we assume that the studied domain has already been differentiated into two regions and the release of signalling molecules is localized in one of those regions~\cite{Rogers:2011:MGG,Shimmi:2003:PPT,Shimmi:2005:FTD}. 
The formation of morphogen gradients then leads to further pattern formation as the cells recognize and interpret the high or low morphogen concentration~\cite{wolpert2015principles}.
Stochastic models helps us to understand the impact of noise (fluctuations) on the pattern~\cite{teimouri2018discrete,kolomeisky2011formation,erban2020stochastic}.

In our model, we distinguish between \textit{signaling molecules} $A$, locally produced, and \textit{morphogen molecules} $B$, which are produced from $A$ through a reaction. Both species move in space via diffusion. The localized production of $A$ (pre-patterning) leads to creation of a morphogen gradient which is then interpreted by cells to produce further patterning~\cite[Section 6.8]{erban2020stochastic}. Since the signaling molecules $A$ are assumed to diffuse at a significantly higher rate compared to $B$, our reaction-diffusion systems are well suited to be analyzed via the mgRDME. Moreover, the primary focus lies on the spatial arrangement of the morphogen $B$, represented by its gradient, while the specific spatial arrangement of signaling molecules $A$ is of less importance. This prioritization is reflected in the use of distinct compartment sizes for the two species, with a finer resolution dedicated to the morphogen $B$.

We study two reaction networks. In Section~\ref{sec2}, we consider a reaction--diffusion system with the two chemical species $A$ (signal) and $B$ (morphogen) being subject to the following three first-order chemical reactions 
\begin{equation}\label{reactions_linear}
\emptyset \longrightarrow A \quad \mbox{(localized)} \qquad \mbox{and}
\qquad
A \longrightarrow B \longrightarrow \emptyset \,,
\end{equation}
where the empty set $\emptyset$ is interpreted as sources and sinks of molecules~\cite{erban2020stochastic}, {\it i.e.}, molecules of signal $A$ are continuously produced and converted into molecules of morphogen~$B$, which are degraded over time. In Section~\ref{sec3}, we replace the first-order conversion reaction $A\to B$ by the second-order \textit{dimerization} reaction $A+A \to B$, {\it i.e.}, the chemical reactions~\eqref{reactions_linear} are changed to
\begin{equation}\label{reactions_dimerization}
\emptyset \longrightarrow A \quad \mbox{(localized)} \qquad \mbox{and} \qquad A+A \longrightarrow B \longrightarrow \emptyset,
\end{equation}
while the diffusion part of the reaction-diffusion model remains the same in both Sections~\ref{sec2} and~\ref{sec3}. Since morphogen gradient systems are effectively one-dimensional (with one `important' direction), we study both systems in a one-dimensional domain. 
For reaction-diffusion systems with first-order chemical kinetics as in~\eqref{reactions_linear}, the one-dimensional results are directly applicable in higher dimensions. 
However, since our second system~(\ref{reactions_dimerization}) includes the second-order reaction (dimerization), the diffusion-limited results will depend on the dimension of the physical space~\cite{Kang:1984:SAK,Montroll:1969:RWL,BenAvraham:1986:KNA}.

For each of the two reaction networks~(\ref{reactions_linear}) and~(\ref{reactions_dimerization}), we formulate the standard RDME, the mgRDME, as well as a `ground truth' model given by particle-based Brownian dynamics. The compartment-based modeling approaches are compared to the Brownian dynamics results based on the differences in their steady-state distributions. Specifically, we define an error function in terms of the distance between the distribution of morphogen molecules $B$ at steady state, and a cost function as a measure of the numerical complexity when simulating the corresponding stochastic processes. Minimizing error and cost over possible choices of compartment sizes for $A$ and $B$ gives rise to a multi-objective optimization problem, which will be studied for both reaction networks. Our results are further summarized in the discussion Section~\ref{sec:conclusion}. 

\section{First-order fast-slow morphogen gradient model}\label{sec2}

In this section, we consider a reaction-diffusion system with the first-order chemical reactions~(\ref{reactions_linear}). We derive analytic expressions for the long-term spatial distribution of particles, examining the standard RDME, the mgRDME and the Brownian dynamics models in Sections~\ref{sec:standard}, \ref{sec:generalized} and~\ref{sec:ground_truth}, respectively. Finally, we compare the models in Section~\ref{sec:comparison} by means of a multi-objective optimization problem.

\subsection{Standard compartment-based model} \label{sec:standard}

\begin{figure}
\centering
\leftline{\hskip 2mm (a)}
\vskip -1mm
\includegraphics[width=0.8\textwidth]{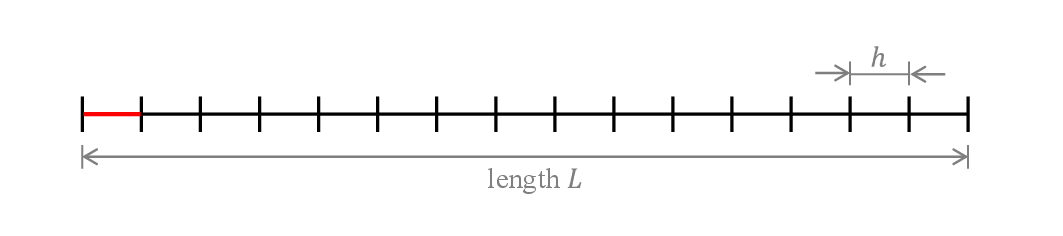}
\leftline{\hskip 2mm (b)}
\vskip -1mm
\includegraphics[width=0.7\textwidth]{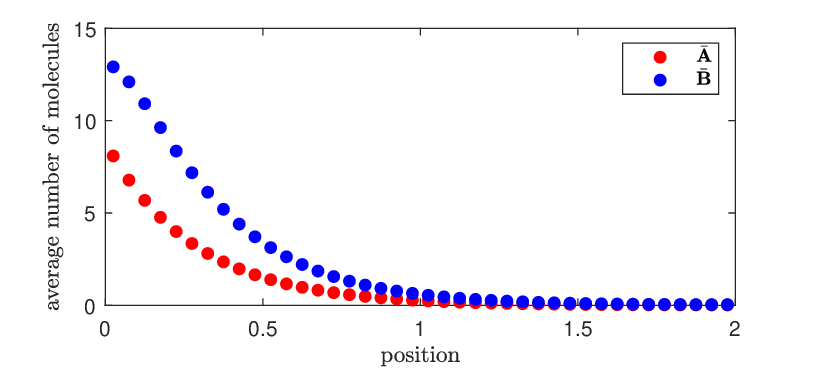}
\vskip 3mm
\caption{{\textbf{Standard compartment-based model.}}
(a) {\it Schematic of the computational domain $[0,L]$ divided into $K$ compartments of size $h$. In the first compartment on the left, highlighted in red, molecules of type $A$ are produced at rate $k_1>0$.} \hfill\break
(b) {\it The average number of molecules of $A$ $($red$)$ and $B$ $($blue$)$ at steady state obtained by solving equation~\eqref{ststequations} for parameters given in~\eqref{parameter_values} and for  $L=2$ and $K=40$ $($i.e., we have $h=L/K=0.05$, $D_A/h^2 = 64$ and $D_B/h^2 = 4)$.
}} 
\label{figure1}
\end{figure}

We consider molecules of chemical species $A$ and $B$ that diffuse in the domain $[0,L]$, where~$L$ is the domain length. The diffusion coefficients are denoted $D_A$ and $D_B$, respectively. They have physical units of [length]$^2$/[time]. As for the standard RDME discretization, we divide the domain $[0,L]$ into $K = L/h$ compartments 
of equal length $h>0$. This domain is schematically shown in Fig.~\ref{figure1}(a).
Denoting the chemical species 
$A$ (resp. $B$) in the $i$-th compartment $[(i-1)h,ih)$ 
by $A_i$ (resp. $B_i$), where $i=1,2,\dots,K$, 
then diffusion corresponds to two chains of ``chemical reactions"~\cite{cao2014stochastic,erban2020stochastic}:
\begin{equation}
A_1
\;
\mathop{\stackrel{\displaystyle\longrightarrow}\longleftarrow}^{d_A}_{d_A}
\;
A_2
\;
\mathop{\stackrel{\displaystyle\longrightarrow}\longleftarrow}^{d_A}_{d_A}
\;
A_3
\;
\mathop{\stackrel{\displaystyle\longrightarrow}\longleftarrow}^{d_A}_{d_A}
\;
\dots
\;
\mathop{\stackrel{\displaystyle\longrightarrow}\longleftarrow}^{d_A}_{d_A}
\;
A_K,
\label{diffA}
\end{equation}
\begin{equation}
B_1
\;
\mathop{\stackrel{\displaystyle\longrightarrow}\longleftarrow}^{d_B}_{d_B}
\;
B_2
\;
\mathop{\stackrel{\displaystyle\longrightarrow}\longleftarrow}^{d_B}_{d_B}
\;
B_3
\;
\mathop{\stackrel{\displaystyle\longrightarrow}\longleftarrow}^{d_B}_{d_B}
\;
\dots
\;
\mathop{\stackrel{\displaystyle\longrightarrow}\longleftarrow}^{d_B}_{d_B}
\;
B_K,
\label{diffB}
\end{equation}
where 
\begin{equation}
d_A = \frac{D_A}{h^2} 
\qquad
\mbox{and}
\qquad 
d_B = \frac{D_B}{h^2}
\end{equation}
are the jump rates. Since $h$ has the physical unit of [length], the jump rates have units [time]$^{-1}$. We assume that molecules of species $A$ are produced in the first compartment on the left,
highlighted in red in Fig.~\ref{figure1}(a). This corresponds to the reaction
\begin{equation}
\emptyset \; \mathop{\longrightarrow}^{k_1/h} \; A_1\,,
\label{creationofA}
\end{equation}
where the rate constant $k_1>0$ has units [time]$^{-1}$.
Scaling with $h$ in reaction~(\ref{creationofA}) is necessary to ensure that after multiplying by the compartment size $h$ -- as is usual for zero-order reactions -- we get a production rate independent of the compartment size. This gives the same production rate in the first compartment for any value of~$h$, and is consistent with our `ground truth' Brownian dynamics model corresponding to partial differential equaitons (PDEs)~\eqref{pde1}--\eqref{pde2} in Section~\ref{sec:ground_truth}. Since~(\ref{reactions_linear}) assumes that $A$ is converted to $B$, and $B$ is degraded in the whole domain, we have the reactions
\begin{equation}
A_i \; \mathop{\longrightarrow}^{k_2}  \; B_i \; \mathop{\longrightarrow}^{k_3 }\;\, \emptyset,
\qquad \mbox{for} \quad i = 1, 2, \dots, K,
\label{conversiondegradation}
\end{equation}
with rate constants $k_2$ and $k_3$ again having physical units of [time]$^{-1}$. The (random) number of particles in the $i$-th compartment at time $t\geq 0$ is denoted by $A_i(t)$ and $B_i(t)$, respectively. Let $p({\mathbf n},{\mathbf m},t)$ be the joint probability that $A_i(t)=n_i$ and
$B_i(t)=m_i$ for $i=1, 2, \dots, K$, where we use the notation 
${\mathbf n}=[n_1,n_2, \dots, n_K] \in \mathbb{N}^K$ and ${\mathbf m} = [m_1,m_2,\dots,m_K] \in \mathbb{N}^K$ to define the system state.
To formulate the RDME corresponding to the chemical reaction
system~\eqref{diffA}--\eqref{conversiondegradation}, we
define operators $\mathcal{O}^+_i, \mathcal{O}^-_i: {\mathbb N}^K \to {\mathbb N}^K$ by
\begin{eqnarray}
\mathcal{O}^+_i [n_1, \dots, n_{i-1}, n_i, n_{i+1}, \dots, n_K] 
&:=& [n_1, \dots, n_{i-1}, n_i + 1, n_{i+1}, \dots,n_K], \qquad
\qquad
\label{Ompdef}
\\
\mathcal{O}^-_i [n_1, \dots, n_{i-1}, n_i, n_{i+1}, \dots, n_K] 
&:=& [n_1, \dots, n_{i-1}, n_i - 1, n_{i+1}, \dots,n_K],
\label{Omidef}
\end{eqnarray}
for $i=1,2,\dots,K$. 
By means of these operators, we define the diffusion operator $\mathcal{D}:L^1\!\left(\mathbb{N}^K\!\!\times\!\mathbb{N}^K\right)\to L^1\!\left(\mathbb{N}^K\!\!\times\!\mathbb{N}^K\right)$, where
\begin{equation}
L^1\!\left(\mathbb{N}^K\!\!\times\!\mathbb{N}^K\right):=\bigg\{f:\mathbb{N}^K\!\!\times\!\mathbb{N}^K \to \mathbb{R} \,\bigg|\, \sum_{{\mathbf n},{\mathbf m}} f({\mathbf n},{\mathbf m}) <\infty\bigg\}
\end{equation}
by
\begin{eqnarray}
    \mathcal{D} f({\mathbf n},{\mathbf m}) & := & 
\frac{D_A}{h^2} 
\sum_{i=1}^{K-1}
\Big\{
(n_i+1)
 \, f (\mathcal{O}^+_i \mathcal{O}^-_{i+1}{\mathbf n},{\mathbf m})
- 
n_i \, f ({\mathbf n},{\mathbf m})
\Big\}
\nonumber
\\
& + &
\frac{D_A}{h^2} 
\sum_{i=2}^{K}
\Big\{
(n_i+1)
 \, f (\mathcal{O}^+_i \mathcal{O}^-_{i-1} {\mathbf n},{\mathbf m})
- 
n_i \, f ({\mathbf n},{\mathbf m})
\Big\} 
\label{diffusion_operator}
\\
& + &
\frac{D_B}{h^2} 
\sum_{i=1}^{K-1}
\Big\{
(m_i+1)
 \, f ({\mathbf n},\mathcal{O}^+_i \mathcal{O}^-_{i+1}{\mathbf m})
- 
m_i \, f ({\mathbf n},{\mathbf m})
\Big\}
\nonumber
\\
& + &
\frac{D_B}{h^2} 
\sum_{i=2}^{K}
\Big\{
(m_i+1)
 \, f ({\mathbf n}, \mathcal{O}^+_i \mathcal{O}^-_{i-1} {\mathbf m})
- 
m_i \, f ({\mathbf n},{\mathbf m})
\Big\} 
\nonumber
\end{eqnarray}
and the reaction operator $\mathcal{R}:L^1\!\left(\mathbb{N}^K\times\mathbb{N}^K\right)\to \mathbb{R}$ by
\begin{eqnarray}
    \mathcal{R} f({\mathbf n},{\mathbf m}) & := & k_1
\Big\{
f (\mathcal{O}^-_1 {\mathbf n},{\mathbf m})
- 
f ({\mathbf n},{\mathbf m})
\Big\}
 \nonumber\\
& + & 
k_2
\sum_{i=1}^{K}
\Big\{
(n_i+1)
 \, f (\mathcal{O}^+_i{\mathbf n},\mathcal{O}^-_i {\mathbf m})
- 
n_i \, f ({\mathbf n},{\mathbf m})
\Big\}
\label{reaction_operator}
\\
& + & 
k_3
\sum_{i=1}^{K}
\Big\{
(m_i+1)
 \, f ({\mathbf n},\mathcal{O}^+_i {\mathbf m})
- 
m_i \, f ({\mathbf n},{\mathbf m})
\Big\}.
\nonumber    
    \end{eqnarray}
Then the reaction--diffusion master equation, 
which corresponds to the system of  reactions 
~\eqref{diffA}--\eqref{conversiondegradation},
can be written as follows
\begin{equation}\label{RDME}
\frac{\partial p}{\partial t}
({\mathbf n},{\mathbf m},t) = (\mathcal{D} + \mathcal{R})
\, p({\mathbf n},{\mathbf m},t)  .
\end{equation}
The stationary distribution is defined by
$$
\phi({\mathbf n},{\mathbf m})
=
\lim_{t \to \infty} 
p ({\mathbf n},{\mathbf m},t)
$$
and satisfies the \textit{stationary reaction--diffusion master equation} 
\begin{equation}\label{statRDME}
0 = (\mathcal{D} + \mathcal{R}) \, \phi({\mathbf n},{\mathbf m}) .
\end{equation}
Solving~\eqref{statRDME}, we obtain the product of Poisson distributions~\cite{jahnke2007solving}
\begin{equation}
\phi({\mathbf n},{\mathbf m})
=
\exp\left[
-
\sum_{i=1}^K
\big(\overline{A}_i + \overline{B}_i\big)
\right]
\prod_{i=1}^K
\frac{\overline{A}_i^{n_i} \overline{B}_i^{\,m_i}}{n_i! \, m_i!} \, ,
\label{statdist}
\end{equation}
where $\overline{A}_i$ and $\overline{B}_i$, for $i=1,2,\dots,K$, satisfy the steady-state equations
\begin{equation}
\left( \frac{D_A}{h^2} \, S - k_2 \, I \right) \overline{\mathbf A} = - k_1  {\mathbf e}_1,
\qquad
\left( \frac{D_B}{h^2} \, S - k_3 \, I \right) \overline{\mathbf B} = - k_2 \overline{\mathbf A}.
\qquad
\label{ststequations}
\end{equation}
Here, $I\in {\mathbb R}^{K \times K}$ is the identity matrix and vectors $\overline{\mathbf A} \in {\mathbb R}^K$,
$\overline{\mathbf B} \in {\mathbb R}^K$, ${\mathbf e}_1 \in {\mathbb R}^K$ and matrix $S \in {\mathbb R}^{K \times K}$ are given by
\begin{equation} \label{ABeS}
\overline{\mathbf A}
=
\left(\!\!
\begin{matrix}
\overline{A}_1 \\
\overline{A}_2 \\
\overline{A}_3 \\
\vdots \\
\overline{A}_{K-1} \\
\overline{A}_K
\end{matrix}
\!\!\right)\!\!,
\;\;
\overline{\mathbf B}
=
\left(\!\!
\begin{matrix}
\overline{B}_1 \\
\overline{B}_2 \\
\overline{B}_3 \\
\vdots \\
\overline{B}_{K-1} \\
\overline{B}_K
\end{matrix}
\!\!\right)\!\!,
\;\;
\mathbf e_1
=
\left(
\begin{matrix}
1 \\
0 \\
0 \\
\vdots \\
0 \\
0
\end{matrix}
\right)\!,
\;\;
S
=
\left(
\begin{matrix}
-1 & 1 & 0 & 0 & \dots & 0 & 0 \\
1 & -2 & 1 & 0 & \dots & 0 & 0 \\
0 & 1 & -2 & 1 & \dots & 0  & 0\\
\vdots & \vdots & \vdots & \vdots & \ddots & \vdots & \vdots \\
0 & 0 & 0 & 0 & \dots & -2 & 1 \\
0 & 0 & 0 & 0 & \dots & 1 & -1
\end{matrix}
\right)\!\!.
\end{equation}
The entries $\overline{A}_i$ and $\overline{B}_i$ are the average numbers of molecules of $A$ and $B$, respectively, in the $i$-th compartment at equilibrium. The solution $(\overline{\mathbf A},\overline{\mathbf B})$ of equation~\eqref{ststequations} is plotted in Figure~\ref{figure1}(b) for the following choice of rate parameter values
\begin{equation}\label{parameter_values}
k_1= 100, \qquad k_2 = 2, \qquad  k_3 = 1, \qquad  D_A=0.16, \qquad D_B=0.01\,,
\end{equation}
showing the morphogen gradient. As the stationary distribution $\phi$ is given by a product of Poisson distributions in~(\ref{statdist}), the values $\overline{A}_i$ and $\overline{B}_i$ not only determine the long-term averages but also the variances of the particle numbers at steady state. Since the values of $\overline{A}_i$ and $\overline{B}_i$ fully characterize the stationary distribution $\phi$, we will compare the models in terms of $\overline{A}_i$ and $\overline{B}_i$.

\subsection{Generalized compartment-based model: mgRDME}\label{sec:generalized}

Assuming that molecules of species $A$ are diffusing significantly faster than those of species $B$ (as is the case of our parameter values~(\ref{parameter_values}) where $D_A \gg D_B$), we choose a larger compartment size for species $A$~\cite{cao2014stochastic}. Let $h_A=1/K_A$ for $K_A\in \mathbb{N}$ be the compartment length for species $A$, while $h_B=1/K_B$ for $K_B\in \mathbb{N}$, $K_B>K_A$, is the compartment length for species $B$. We choose the values $K_A$ and $K_B$ such that the ratio $\gamma:=K_B/K_A=h_A/h_B >1$ is a natural number, $\gamma \in \mathbb{N}$. In the mgRDME formulation~\cite{cao2014stochastic}, reaction \eqref{creationofA} is replaced by
\begin{equation}
\emptyset \; \mathop{\longrightarrow}^{k_1/h_A} \;\, A_1,
\label{creationofA_generalized}
\end{equation}
which implies again that $A$-particles are produced at rate $k_1$ in the first compartment on the left (of the coarser $A$-discretization), independently of the grid size $h_A$. 
The reaction $A_i \; \longrightarrow  \; B_i$ in 
\eqref{conversiondegradation} is generalized to
\begin{equation}\label{A_j_B_i}
A_j \; \stackrel{k_2/\gamma}{\longrightarrow}  \; B_i, \quad j=1,2,\dots,K_A, \quad \mbox{for} \;\; i \in \mathcal{I}(j) := \{ (j-1)\gamma +1,\, \dots \,, \,j\gamma\},
\end{equation}
which means that after conversion from $A$ to $B$ the resulting $B$-molecule is placed uniformly in one of the $\gamma$ (smaller) compartments that overlap with the $j$-th (large) compartment of the reacting $A$-particle. The diffusion operator~\eqref{diffusion_operator} is generalized to 
\begin{eqnarray}
\tilde{\mathcal{D}}f({\mathbf n},{\mathbf m})
& := & 
\frac{D_A}{h_A^2} 
\sum_{j=1}^{K_A-1}
\Big\{
(n_j+1)
 \, f (\mathcal{O}^+_j \mathcal{O}^-_{j+1}{\mathbf n},{\mathbf m})
- 
n_j \, f ({\mathbf n},{\mathbf m})
\Big\}
\nonumber
\\
& + &
\frac{D_A}{h_A^2} 
\sum_{j=2}^{K_A}
\Big\{
(n_j+1)
 \, f (\mathcal{O}^+_j \mathcal{O}^-_{j-1} {\mathbf n},{\mathbf m})
- 
n_j \, f ({\mathbf n},{\mathbf m})
\Big\} 
\label{diffusion_operator_generalized}
\\
& + &
\frac{D_B}{h_B^2} 
\sum_{i=1}^{K_B-1}
\Big\{
(m_i+1)
 \, f ({\mathbf n},\mathcal{O}^+_i \mathcal{O}^-_{i+1}{\mathbf m})
- 
m_i \, f ({\mathbf n},{\mathbf m})
\Big\}
\nonumber
\\
& + &
\frac{D_B}{h_B^2} 
\sum_{i=2}^{K_B}
\Big\{
(m_i+1)
 \, f ({\mathbf n}, \mathcal{O}^+_i \mathcal{O}^-_{i-1} {\mathbf m})
- 
m_i \, f ({\mathbf n},{\mathbf m})
\Big\} ,
\nonumber
\end{eqnarray}
while the reaction operator~\eqref{reaction_operator} now reads 
\begin{eqnarray}
\tilde{\mathcal{R}}f({\mathbf n},{\mathbf m})
& := & 
k_1
\Big\{
f (\mathcal{O}^-_1 {\mathbf n},{\mathbf m})
- 
f ({\mathbf n},{\mathbf m})
\Big\} \nonumber
\\
& + &
\frac{k_2}{\gamma}
\sum_{j=1}^{K_A}\sum_{i\in \mathcal{I}(j)}
\Big\{
(n_j+1)
 \, f (\mathcal{O}^+_j{\mathbf n},\mathcal{O}^-_i {\mathbf m})
- 
n_j \, f ({\mathbf n},{\mathbf m})
\Big\} \qquad
\label{reaction_operator_generalized}
\\
& + &
k_3
\sum_{i=1}^{K_B}
\Big\{
(m_i+1)
 \, f ({\mathbf n},\mathcal{O}^+_i {\mathbf m})
- 
m_i \, f ({\mathbf n},{\mathbf m})
\Big\}
\nonumber
.
\end{eqnarray}
Dividing by $\gamma$ in the second line of~\eqref{reaction_operator_generalized} goes in line with summing over the set $\mathcal{I}(j)$ defined in~\eqref{A_j_B_i} which contains $|\mathcal{I}(j)|=\gamma$ elements; this can be interpreted as a sum of $\gamma$ reactions $A_j \to B_i$ that share the rate $k_2$. In total, we obtain the \textit{multi-grid reaction--diffusion master equation} (mgRDME)
\begin{equation}\label{RDME_multi}
\frac{\partial p}{\partial t}({\mathbf n},{\mathbf m},t) \,=\, (\tilde{\mathcal{D}} + \tilde{\mathcal{R}})
\,p({\mathbf n},{\mathbf m},t)  .
\end{equation}
Solving the corresponding stationary mgRDME, given by $0=(\tilde{\mathcal{D}} + \tilde{\mathcal{R}})\,\phi({\mathbf n},{\mathbf m})$ in analogy to~\eqref{statRDME}, we obtain a product of Poisson distributions similar to~\eqref{statdist}:
\begin{equation}
\phi({\mathbf n},{\mathbf m})
=
\exp\left[
-
\sum_{j=1}^{K_A}
\overline{A}_j - \sum_{i=1}^{K_B} \overline{B}_i
\right]
\,
\prod_{j=1}^{K_A}
\frac{\overline{A}_j^{n_j}}{n_j!} 
\,
\prod_{i=1}^{K_B}\frac{ \overline{B}_i^{\,m_i}}{m_i!},
\label{statdist2}
\end{equation}
where $\overline{A}_j$ and $\overline{B}_i$, for $j=1,2,\dots,K_A$, $i=1,2,\dots,K_B$, satisfy the generalized steady-state equations
\begin{equation}
\left( \frac{D_A}{h_A^2} \, S_A - k_2 \, I_A \right) \overline{\mathbf A} = - k_1 h_A^3 {\mathbf e}_1,
\qquad
\left( \frac{D_B}{h_B^2} \, S_B - k_3 \, I_B \right) \overline{\mathbf B} = - \frac{k_2}{\gamma} M\overline{\mathbf A}.
\label{ststequations2}
\end{equation}
Here, $I_A\in {\mathbb R}^{K_A \times K_A}$ and  $I_B\in {\mathbb R}^{K_B \times K_B}$ are identity matrices and vectors $\overline{\mathbf A} \in {\mathbb R}^{K_A}$,
$\overline{\mathbf B} \in {\mathbb R}^{K_B}$, ${\mathbf e}_1 \in {\mathbb R}^{K_A}$ and matrices $S_A \in {\mathbb R}^{K_A \times K_A}$, $S_B \in {\mathbb R}^{K_B \times K_B}$ are defined in an analogous manner as before in~\eqref{ABeS}. In addition, there is the block matrix $M\in \mathbb{R}^{K_B \times K_A}$ given by
\begin{equation*}
    M= \left(\begin{matrix}M_1 \\
    \vdots \\
    M_{K_A}
    \end{matrix}\right),\quad M_j = \left(\begin{matrix}
    0 & \dots & 0 & 1 & 0 & \dots & 0  \\
    \vdots &  & \vdots & \vdots & \vdots &  & \vdots \\
    0 & \dots & 0 & 1 & 0 & \dots & 0 \end{matrix}\right)\in \mathbb{R}^{\gamma \times K_A}, \quad j = 1,\dots,K_A,
\end{equation*}
with the non-zero entries ({\it i.e.}, the ones) in the block $M_j\in \mathbb{R}^{\gamma \times K_A}$ placed in column $j$. 
As in the standard setting, $\overline{A}_j$ and $\overline{B}_i$ give both the long-term mean and variance of the population sizes in the respective boxes, and they fully characterize the stationary distribution $\phi$. 

\begin{figure}
\centering
\begin{subfigure}[b]{\textwidth}
\includegraphics[width=0.49\textwidth]{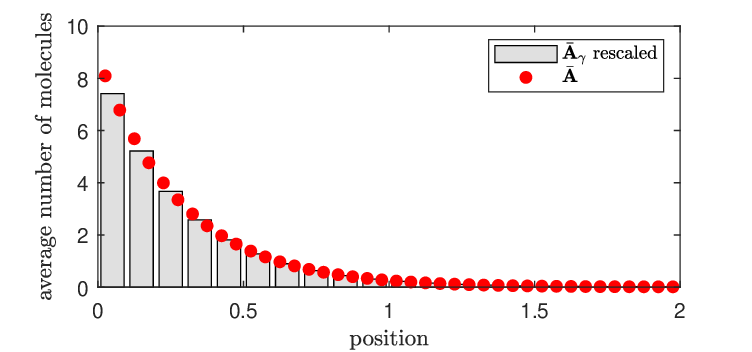}
\hfill
\includegraphics[width=0.49\textwidth]{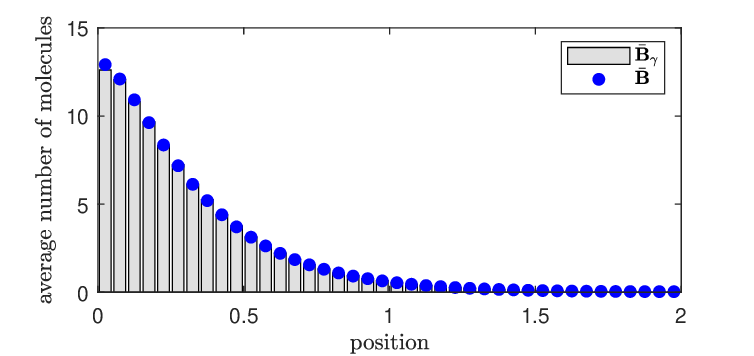}\caption{$\gamma=2$}
\end{subfigure}
\begin{subfigure}[b]{\textwidth}\includegraphics[width=0.49\textwidth]{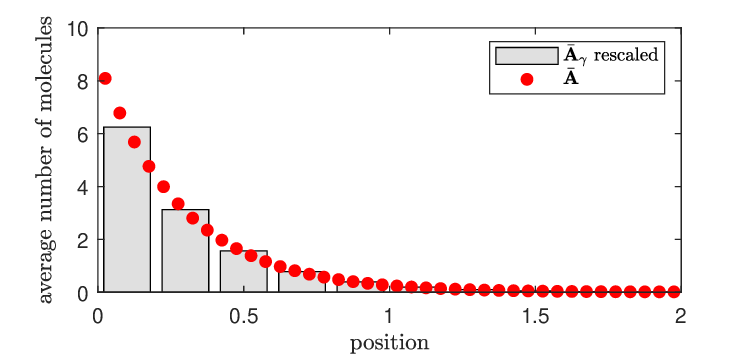}
\hfill
\includegraphics[width=0.49\textwidth]{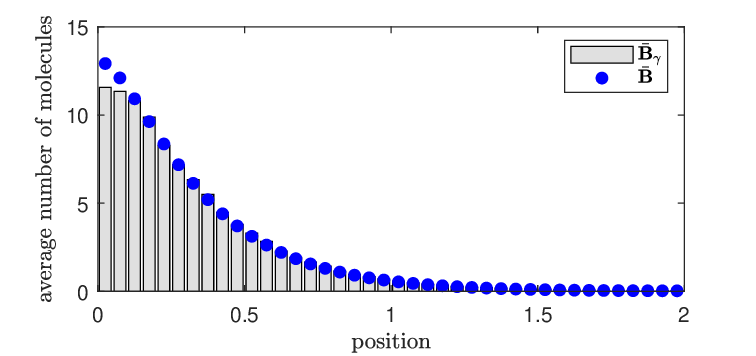}
\caption{$\gamma=4$}
\end{subfigure}
\caption{\textbf{Standard RDME vs. mgRDME.} {\it Average number of molecules of $A$ and $B$ at equilibrium obtained by solving the standard stationary RDME~\eqref{ststequations} $($red/blue dots$)$ and by solving the stationary mgRDME~\eqref{ststequations2} $($grey bars$)$ for rate constants given in~\eqref{parameter_values} and for $L=2$, $K_B=40$ and} (a) $\gamma = 2$, (b) $\gamma = 4$. }
\label{fig:average_gerenalized}
\end{figure}

\paragraph{Model comparison: standard {\rm RDME} versus {\rm mgRDME}.} 
In the following, we denote the solution of the generalized steady-state equation \eqref{ststequations2} by $\bar{\mathbf A}_\gamma=(\bar{A}^{(\gamma)}_j)_{j=1,2,\dots,K_A}$, $\bar{\mathbf B}_\gamma=(\bar{B}^{(\gamma)}_i)_{i=1,2,\dots,K_B}$ in order to emphasize its dependence on the ratio $\gamma=K_B/K_A$ and to distinguish from the solution $\bar{\mathbf A}=\bar{\mathbf A}_1,$ $\bar{\mathbf B}=\bar{\mathbf B}_1$ of the standard compartment-based model given by equation~\eqref{ststequations}. Figure~\ref{fig:average_gerenalized} shows the solution $\bar{\mathbf A}_\gamma,\bar{\mathbf B}_\gamma$ of equation~\eqref{ststequations2} for the same parameter values as in Figure~\ref{figure1}(b) and for $\gamma=2$ and $\gamma=4$. Instead of $\bar{\mathbf A}_\gamma$ we plot the rescaled values $\tilde{A}^{(\gamma)}_i:=\bar{A}^{(\gamma)}_j/\gamma$ for $i\in \mathcal{I}(j)$. This is for the purpose of comparability with the solution $\bar{\mathbf A}$ of the standard steady-state equation~\eqref{ststequations} plotted in Figure~\ref{figure1}(b). 

\begin{figure}
\centering
\includegraphics[width=0.5\textwidth]{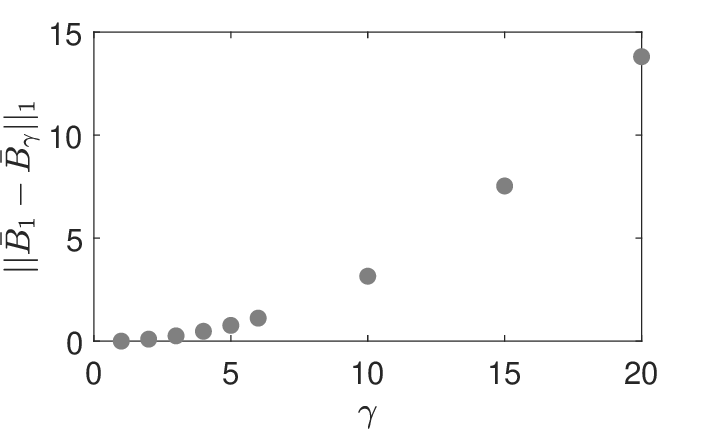}
\caption{{\textbf{Difference between standard RDME and mgRDME.}} {\it Difference $\|\bar{\mathbf B}_1-\bar{\mathbf B}_\gamma\|_1$ as a function of the ratio $\gamma=K_B/K_A$ for rate constants given in~\eqref{parameter_values} and for $L=2$,  $K_B=60$ and $K_A\in\{60,30,20,15,12,10,6,4,3\}$, where $\bar{\mathbf B}_1$ the solution of the standard steady-state equation \eqref{ststequations},  and $\bar{\mathbf B}_\gamma$ the solution of the generalized steady-state equation \eqref{ststequations2}.}}\label{fig:error_Bbar}
\end{figure}

We observe a qualitatively good agreement of the generalized solutions with the original one. To see how this agreement depends on the level of coarsening, we plot the difference  $\|\bar{\mathbf B}_1-\bar{\mathbf B}_\gamma\|_1$ as a function of $\gamma=K_B/K_A$ in Figure~\ref{fig:error_Bbar}. We see that the difference monotonically increases with $\gamma$, which is to be expected since large values of $\gamma$ represent a high degree of coarsening. However, the increase of the error is nonlinear, and there is an area of values ($\gamma \lesssim 5 $) where the error is sufficiently small to be deemed acceptable. Before we study the approximation quality in more detail, we define in the following section the `ground truth' model given by particle-based Brownian dynamics. 

\subsection{`Ground truth' particle-based model: Brownian dynamics} \label{sec:ground_truth}

As our `ground truth' model, we will consider a model with two species $A$ and $B$ diffusing and reacting in the unbounded domain~${\mathbb R}$~\cite{erban2020stochastic}. Molecules of $A$ are released at the origin, $x=0,$ with rate $2 k_1$, which has units [sec$^{-1}$].  Due to symmetry, this is equivalent to studying the same model on the half-line $[0,\infty)$ with the rate constant~$k_1$ and reflecting boundary conditions at $x=0$~\cite{Erban:2007:RBC}. Molecules of $A$ and $B$ are diffusing with diffusion constants $D_A$ and $D_B$, respectively. They are subject to the conversion and degradation reactions 
\begin{equation}
A \; \mathop{\longrightarrow}^{k_2}  \; B \; \mathop{\longrightarrow}^{k_3 }\;\, \emptyset.
\label{conversiondegradationBD}
\end{equation}
 We compare the Brownian dynamics to the simulation on the finite interval~$[0,L]$, which is discretized into compartments in the RDME and mgRDME frameworks. The~effect of the boundary at $x=L$ is discussed in Section~\ref{secboundBD} of the Appendix.

Let $a(x,t)$ and $b(x,t)$ be the concentrations of molecules of $A$ and $B$, respectively, that is $a(x,t) \, \mbox{d}x$ denotes the average number of molecules in the interval $[x,x+\mbox{d}x)$ at time $t$ in the Brownian dynamics model. Then $a(x,t)$ and $b(x,t)$ satisfy the following PDEs~\cite[equation (1)]{kicheva2007kinetics}
\begin{eqnarray}
\frac{\partial a}{\partial t} 
&=&
D_A 
\frac{\partial^2 a}{\partial x^2} 
- k_2 \, a + 2 k_1 \, \delta(x), 
\label{pde1}
\\
\frac{\partial b}{\partial t} 
&=&
D_B 
\frac{\partial^2 b}{\partial x^2} 
+ k_2 \, a - k_3 \, b,
\label{pde2}
\end{eqnarray}
where $x \in {\mathbb R}$, and $\delta(x)$ is the Dirac delta function. We consider the boundary conditions
\begin{equation}
\lim_{x \to \pm \infty}
a(x,t)
=
\lim_{x \to \pm \infty}
b(x,t)
=
0.
\label{boundcond}    
\end{equation}Let $\bar{a}(x)$ and $\bar{b}(x)$ be the corresponding stationary distributions, {\it i.e.}
$$
\bar{a}(x)
=
\lim_{t \to \infty} 
a(x,t),
\qquad
\bar{b}(x)
=
\lim_{t \to \infty} 
b(x,t).
$$
They satisfy the steady-state equations
\begin{eqnarray}
0 
&=&
D_A 
\frac{\mbox{d}^2 \bar{a}}{\mbox{d} x^2} 
- k_2 \, \bar{a} + 2 k_1 \, \delta(x) \, , 
\label{ststeq1}
\\
0
&=&
D_B 
\frac{\mbox{d}^2 \bar{b}}{\mbox{d} x^2} 
+ k_2 \, \bar{a} - k_3 \, \bar{b} \, ,
\label{ststeq2}
\end{eqnarray}
with boundary conditions
\begin{equation}
\lim_{x \to \pm \infty}
\bar{a}(x)
=
\lim_{x \to \pm \infty}
\bar{b}(x)
=
0.
\label{boundcondODE}   
\end{equation}
We notice that the PDE~\eqref{pde1} (resp. the ODE~\eqref{ststeq1}) does not depend on $b(x,t)$ (resp. $\bar{b}(x)$). In what follows, we will therefore first solve the equation~(\ref{pde1}) for the chemical species~$A$, and then we will substitute it into the second equation~(\ref{pde2}) for the chemical species $B$ to get our `ground truth' solution. In the following, we will do this for the steady-state problem. The time-dependent problem is considered in the appendix, see Section~\ref{sectimeBD}. 

\paragraph{Steady-state solution}

Since the second derivative of the absolute value function, $|x|$, is equal to $2 \delta(x)$, we can write the solution to equation~\eqref{ststeq1} with boundary conditions~(\ref{boundcondODE}) as
\begin{equation}
\bar{a}(x) 
= 
\frac{k_1}{\sqrt{D_A \, k_2}} \,
\exp \left[ 
-\sqrt{\frac{k_2}{D_A}} \, |x|
\right].
\label{assteadystatesol}
\end{equation}
This result is also obtained when taking the limit $t\to \infty$ of the time-dependent solution $a(x,t)$ as given in equation~\eqref{solaxt} of Appendix~\ref{sectimeBD}. 
Substituting into equation~(\ref{ststeq2}), we get
$$
D_B 
\frac{\mbox{d}^2 \bar{b}}{\mbox{d} x^2} 
+ k_1
\sqrt{\frac{k_2}{D_A}} \,
\exp \left[ 
-\sqrt{\frac{k_2}{D_A}} \, |x|
\right]
- k_3 \, \bar{b}
=
0.
$$
Consequently, using boundary conditions~(\ref{boundcondODE}), we obtain
\begin{equation}
\bar{b}(x)
=
\frac{k_1 \sqrt{k_2 \, D_A}}{D_A \, k_3  - D_B \, k_2}
\,
\exp \left[ 
-
\sqrt{\frac{k_2}{D_A}} \, |x|
\right]
+
\frac{k_1 k_2 \sqrt{D_B}}{(D_B \, k_2 - D_A \, k_3) \sqrt{k_3}}
\,
\exp \left[ 
- \sqrt{\frac{k_3}{D_B}} \, |x|
\right]
\label{bssteadystatesol}
\end{equation}
for the concentration $\bar{b}$ of molecuels of $B$ at equilibrium. 
Our compartment-based models simulate the problem on the finite interval $[0,L]$, with zero-flux boundary condition at $x=L.$ To get the `ground truth' solution~\eqref{bssteadystatesol}, we have used the infinite domain, ${\mathbb R}$, with boundary conditions~(\ref{boundcondODE}). In Section~\ref{secboundBD} we show that the effect of non-flux boundary conditions on the `ground truth' solution $\bar{a}$ vanishes for $L\to \infty$. In particular, the concentration is negligible at $x\geq L$ as long as $L$ is large enough, and we can use~\eqref{assteadystatesol} and \eqref{bssteadystatesol} for a comparison with the compartment-based models. 

\subsection{Model comparison}\label{sec:comparison}

Since the reaction system under consideration is a first-order reaction network, the~standard compartment-based model of Section~\ref{sec:standard} converges to the Brownian dynamics solution for $h\to 0$~\cite{engblom2009simulation,erban2020stochastic}. Our goal is now to compare the Brownian dynamics to the compartment-based model for finite $h$, both for the standard RDME and for the mgRDME. In application, one is mainly interested in the output given by the product $B$, which motivates to define the distance between the models by means of the average number of molecules of $B$. 

\subsubsection{Error definition}

Given the steady-state solution $\bar{\mathbf B}_h=(\bar{B}_1,\bar{B_2},\dots,\bar{B}_{K_B})$ of the mgRDME model with $K_B$ boxes of size $h=L/K_B$ (which agrees with the standard RDME solution when choosing $\gamma=1$), we define the piecewise constant function $\bar{b}_h:[0,L]\to [0,\infty)$,
\begin{equation}\label{bar_b_h}
    \bar{b}_h(x) := \bar{B}_i/h \quad \mbox{ for } x\in [(i-1)h,ih), \quad i=1,2,\dots,K_B,
\end{equation} 
which approximates the steady-state concentration of molecules of $B$ as it depends on the location $x$. The $L^1$-distance to the Brownian dynamics solution $\bar{b}$ defines
\begin{equation}\label{err1}
    \err(K_B,\gamma) := ||\bar{b}-\bar{b}_h ||_{L^1} = \int_0^L |\bar{b}(x)-\bar{b}_h(x)| \, \mbox{d}x = \sum_{i=1}^{K_B} \int_{(i-1)h}^{ih} \!\!\!\!|\bar{b}(x)-\bar{B}_i| \, \mbox{d}x\,, \qquad
\end{equation}
the spatial error in dependence on $K_B$ and on the ratio $\gamma=K_B/K_A$. We note that this error would be positive even if $\bar{b}_h$ was a discretization of $\bar{b}$ (and not the solution of the compartment-based model), simply because of the coarse-graining. We thus expect the error to naturally decrease with decreasing $h$. 

\begin{figure}
\centering
\includegraphics[width=0.6\textwidth]{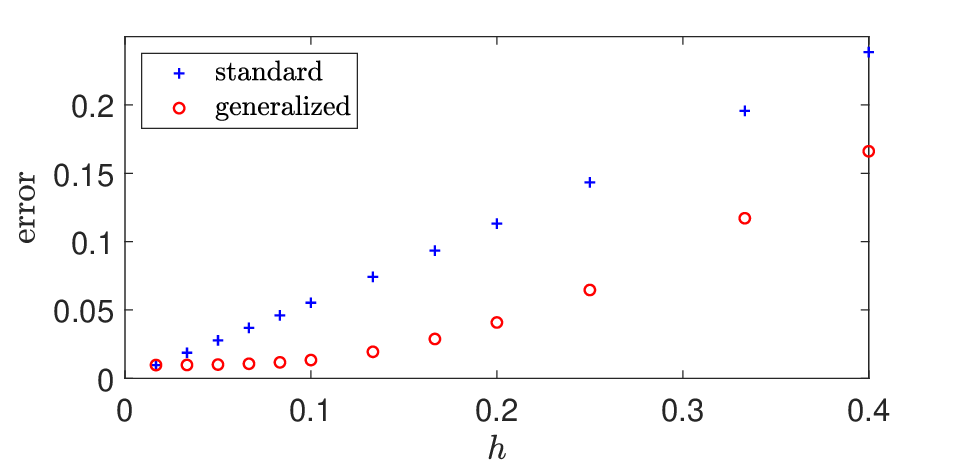}
\caption{{\textbf{Error in dependence on the compartment size.}} {\it Error $\err(K_B,\gamma)$, defined in \eqref{err1}, between the steady-state solution  $\bar{b}$ of the `ground truth' model given by \eqref{bssteadystatesol} and its approximation by the standard and the generalized compartment based models with solutions $\bar{\mathbf B}_h$ in dependence on the compartment size $h$. We have $h_A=h_B=h$ for the standard model, while $h_A=h$ and $h_B=h^*=1/60$ for the multi-grid model. Rate constants given in~\eqref{parameter_values} and $L=2$.}}
\label{fig:error_B}
\end{figure}

Figure~\ref{fig:error_B} shows this error in dependence on the grid size $h$, where $h_A=h_B=h$ for the standard RDME, while $h_A=h$ and fixed $h_B=h^*=1/60$ for the generalized mgRDME. For both scenarios, there is a monotone increase of the error with $h$, which is to be expected in a first-order reaction system. The generalized model shows better results as compared to the standard one, which is due to the fixed small grid size $h^*$ for $B$. 
However, the error $\err(K_B,\gamma)$ seems to be linear in $h$ for $\gamma=1$ (standard model, blue crosses), while it is nonlinear in $h=h_A$ for the generalized system and stays close to zero for small $h=h_A$. We can say that for $h=h_A \leq 0.1$ ($\gamma \leq 6$) the generalized mgRDME model gives a good approximation, while the standard compartment-based model already shows a clear error. 

Despite the availability of an analytical solution for the steady-state distribution, there remains a fundamental interest in stochastic simulations of the dynamics, which allow for a more comprehensive exploration of the system's behavior including complex interactions and rare events.
We point out that, in Figure~\ref{fig:error_B} and for a given $h$, the numerical effort for simulating trajectories is larger for the multi-grid model compared to the standard model because due to the retention of the small grid size $h_B=h^*$ for $B$. However, the overall goal of the generalized method is to reduce numerical cost while simultaneously minimizing errors. Therefore, determining an optimal value for~$\gamma$ entails multi-objective optimization, a task we will address next.

\subsubsection{Multi-objective optimization}
\label{sec:multi-obj}

In this section, we study the dependence of the spatial error $\err(K_B,\gamma)$ defined in~\eqref{err1} and of the numerical effort on the grid sizes $h_A$ and $h_B$. We calculate error and cost for different parameter values to show the advantage of the generalized method over the standard method. To define the cost function, we use propensities at steady state because these indicate the runtime of our numerical simulations. 

\paragraph{Propensities at steady state}

The total number of particles at steady state is given by\footnote{The values result from setting the related reaction rate equations to zero, {\it i.e.}, from  solving $k_1-k_2\bar{A}_{\text{total}}=0$ and $k_2\bar{A}_{\text{total}}-k_3\bar{B}_{\text{total}}=0$. This works only for first-order reaction systems. }
\begin{equation} \label{eq:AB_total}
\bar{A}_{\text{total}}= \sum_{i=1}^{K_A} \bar{A}_i= \frac{k_1}{k_2}, \qquad\quad  \bar{B}_{\text{total}}= \sum_{i=1}^{K_B} \bar{B}_i=\frac{k_2}{k_3}\bar{A}_{\text{total}} = \frac{k_1}{k_3}.\qquad
\end{equation}
The overall propensity for a reaction to take place at steady state (no matter which reaction or where in space) is
\begin{equation} \label{prop_react}
\bar{r}_{\text{reac}} := k_1 + k_2 \, \bar{A}_{\text{total}} + k_3 \,\bar{B}_{\text{total}} =3\,k_1\,.
\end{equation}
Analogously, the overall propensity for a diffusive jump to take place at steady state is
\begin{align}
\bar{r}_{\text{diff}} & := \frac{D_A}{h_A^2}(2 \bar{A}_{\text{total}} - \bar{A}_1-\bar{A}_{K_A}) + \frac{D_B}{h_B^2} (2\bar{B}_{\text{total}} -\bar{B}_1-\bar{B}_{K_B})\nonumber \\
&\lessapprox  2\frac{D_A}{h_A^2}\bar{A}_{\text{total}} +2\frac{D_B}{h_B^2}  \bar{B}_{\text{total}} \;
= \; 2\frac{D_A}{h_A^2}\frac{k_1}{k_2} +2\frac{D_B}{h_B^2}  \frac{k_1}{k_3} \label{prop_diff}
\end{align}
with the negative terms in the first line resulting from the fact that diffusive jumps at the outer boundaries of the domain can only go in one direction. The latter expression is a good approximation as long as the population sizes in the outer boxes are not too large (as compared to the total population size), which is the case in the scenarios considered.

\begin{figure}
 \centering
\includegraphics[width=1\textwidth]{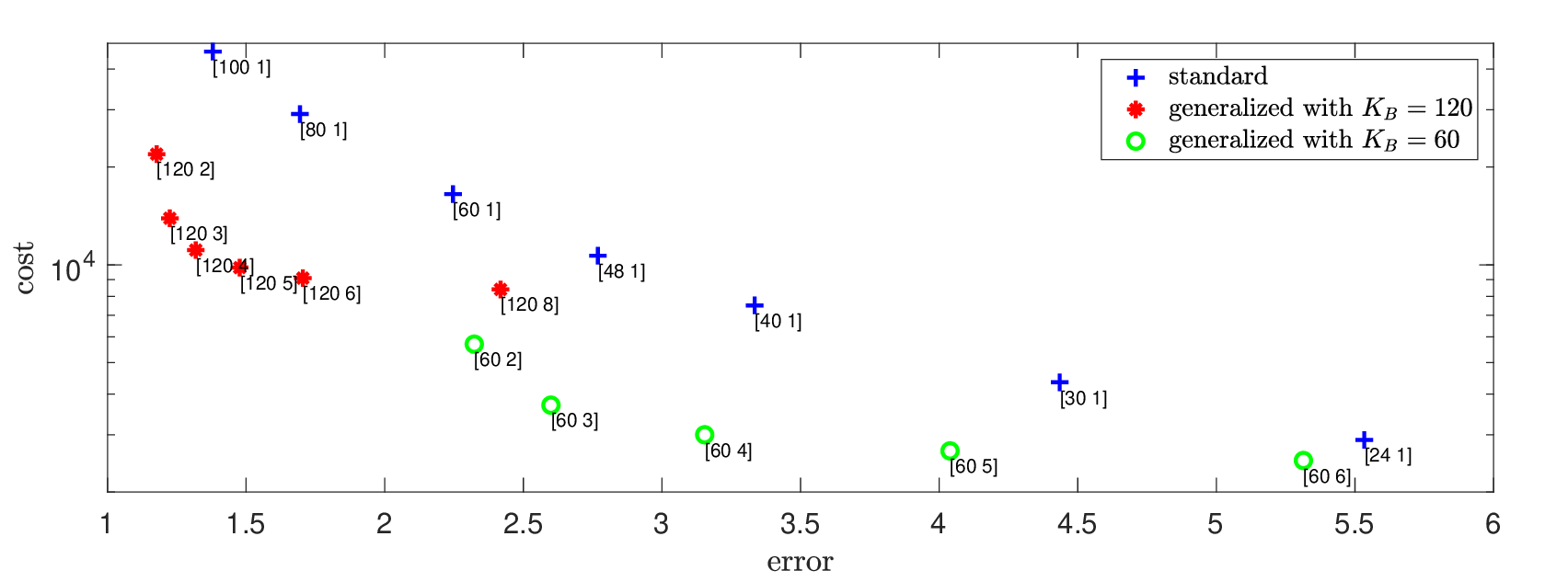}
\caption{{\textbf{Error and cost depending on the method.}} {\it Error $\err(K_B,\gamma)$ as defined in~\eqref{err1} and numerical cost $c(K_B,\gamma)$ as defined in~\eqref{cost} for different values of $K_B$ and $\gamma=K_B/K_A$ $($given by labels of the form $[K_B\; \gamma])$. The colored symbols are for orientation: blue pluses for the standard RDME, red stars for the generalized mgRDME with $K_B=120$ and green circles for the generalized mgRDME with $K_B=60$. \hfill\break
Parameters are given in~\eqref{parameter_values} and $L=2$. 
}}
\label{fig:MOO}
\end{figure}

The number of iterations steps needed to create a trajectory of the dynamics (starting in steady state, using the Gillespie algorithm~\cite{gillespie1977exact,gillespie2007stochastic}) scales with $\bar{r}_{\text{reac}}+\bar{r}_{\text{diff}}$. This motivates to define the cost function as a sum of~\eqref{prop_react}~and~\eqref{prop_diff}, giving 
\begin{equation} 
\label{cost}
c(K_B,\gamma) := 3 \, k_1 + 2 \, \frac{D_A}{h_A^2} \frac{k_1}{k_2} + 2\, \frac{D_B}{h_B^2}  \frac{k_1}{k_3},
\end{equation}
where $h_B=L/K_B$ and $h_A=\gamma h_B$. Figure~\ref{fig:MOO} shows the values of error and cost depending on $K_B$ and $\gamma$ (where $\gamma=1$ corresponds to the standard RDME model). We see that the generalized model clearly outperforms the standard model: 
The Pareto front exclusively comprises of points derived from the generalized model. Consequently, the decrease in spatial resolution for species $A$ introduced by the generalized model minimally impacts the error in the $B$-solution, relative to the alteration in numerical cost resulting from this coarse-graining. We note the special role of $\gamma=4=\sqrt{D_A/D_B}$, where the jump rates of the two species coincide. 

\medskip

\begin{remark} We note that the scenario would change if we were to define an error metric that incorporates the distance within the $A$-population.  This is evident because $||\mathbf{A}-\bar{\mathbf{A}}(K_B,\gamma)||_1$ unmistakably rises with $\gamma$, irrespective of $K_B$. \end{remark}

\medskip

\noindent
In the examined first-order reaction system, only the production of $A$ depends on location, whereas the other reactions occur independently of the molecules' spatial position. A~stronger effect of the grid size on the reaction dynamics is expected for a second-order reaction system, which is studied next.

\section{Fast-slow morphogen gradient with dimerization} \label{sec3}

We consider the fast-slow morphogen gradient system of Section~\ref{sec2}, replacing the conversion reaction $A\to B$ by the second-order reaction of \textit{dimerization} $A+A \to B$, {\it i.e.}, the chemical reactions~\eqref{reactions_linear} are changed to
(\ref{reactions_dimerization}), while the diffusion part of the reaction-diffusion model remains consistent with that described in the previous section. Again, we choose the one-dimensional domain $[0,L]$. The associated modeling approaches are introduced in Section~\ref{sec:models_dim} and compared in Section~\ref{sec:comparison_dim}. 

\subsection{Modeling approaches for the dimerization system} \label{sec:models_dim}

Considering the compartment-based models, the diffusion operators are the same as in Section~\ref{sec2}, {\it i.e.}, for the standard RDME model it is given by $\mathcal{D}$ as defined in \eqref{diffusion_operator}, and for the generalized mgRDME model it is given by $\tilde{\mathcal{D}}$ as defined in \eqref{diffusion_operator_generalized}. In contrast, the reaction operators now explicitly depend on the grid size $h$ via the second-order reaction $A+A\to B$. For the standard compartment-based model, given by the RDME $\frac{\partial p}{\partial t}({\mathbf n},{\mathbf m},t) = (\mathcal{D} + \mathcal{R}) \, p({\mathbf n},{\mathbf m},t) $, it reads 
\begin{eqnarray}
&&\mathcal{R} f({\mathbf n},{\mathbf m}) \, := \, k_1
\Big\{
f (\mathcal{O}^-_1 {\mathbf n},{\mathbf m})
- 
f ({\mathbf n},{\mathbf m})
\Big\}
 \nonumber\\
\qquad & + & 
\frac{k_2}{h}
\sum_{i=1}^{K}
\Big\{
(n_i+2)(n_i+1)
 \, f (\mathcal{O}^+_i\mathcal{O}^+_i{\mathbf n},\mathcal{O}^-_i {\mathbf m})
- 
\max\{n_i(n_i-1),0\} \, f ({\mathbf n},{\mathbf m})
\Big\}
\nonumber
\\
\qquad & + & 
k_3
\sum_{i=1}^{K}
\Big\{
(m_i+1)
 \, f ({\mathbf n},\mathcal{O}^+_i {\mathbf m})
- 
m_i \, f ({\mathbf n},{\mathbf m})
\Big\}
\nonumber    
\end{eqnarray}
for  $h=h_A=h_B$. The rate constant $k_2$ now has physical units [length]/[time], {\it i.e.}, units of the `one-dimensional volume' divided by time, because $k_2$ is the rate constant of a bimolecular reaction, in contrast to $k_2$ in \eqref{reaction_operator}, which has physical units [time]$^{-1}$. Dividing by $h$ in the second line stems from the standard scaling of a second-order reaction rate by the volume of the domain in which it takes place, which is given by~$h$ for the one-dimensional case under consideration. 
Similarly, for the generalized mgRDME model, we have 
\begin{eqnarray}
&&\tilde{\mathcal{R}}f({\mathbf n},{\mathbf m})
\, := \, 
k_1
\Big\{
f (\mathcal{O}^-_1 {\mathbf n},{\mathbf m})
- 
f ({\mathbf n},{\mathbf m})
\Big\} \nonumber
\\
&& \;\; + 
\frac{k_2}{h_A\gamma}
\sum_{j=1}^{K_A}\sum_{i\in \mathcal{I}(j)}
\Big\{
(n_j+2)(n_j+1)
 \, f (\mathcal{O}^+_j\mathcal{O}^+_j{\mathbf n},\mathcal{O}^-_i {\mathbf m})
- 
\max \{n_j(n_j-1)\} \, f ({\mathbf n},{\mathbf m})
\Big\}
\nonumber
\\
&& \;\;
+
k_3
\sum_{i=1}^{K_B}
\Big\{
(m_i+1)
 \, f ({\mathbf n},\mathcal{O}^+_i {\mathbf m})
- 
m_i \, f ({\mathbf n},{\mathbf m})
\Big\}
\nonumber
.
\end{eqnarray}
The factor $1/\gamma$ in the second summand results from splitting the reaction $A+A\to B$ occurring in the $j$-th compartment into $|\mathcal{I}(j)|=\gamma$ possible ones, depending on the placement of the product $B$:
\begin{equation}
A_j+A_j \; \stackrel{k_2/\gamma}{\longrightarrow}  \; B_i, \quad j=1,2,\dots,K_A, \quad  i \in \mathcal{I}(j) := \{ (j-1)\gamma +1,\, \dots \,,j\gamma\},
\end{equation}
which conforms with the first-order case, see~\eqref{A_j_B_i} and \eqref{reaction_operator_generalized}.

\paragraph{'Ground truth' particle-based model: Brownian dynamics} 

As the 'ground truth' we choose particle-based dynamics given by the Doi or $\lambda$-$\varrho$ model~\cite{doi1976stochastic,doi1976second,erban2009stochastic,Lipkova:2011:ABD}. Particles move in space by Brownian motion, and two particles of species $A$ undergo dimerization $A+A \to B$ at rate $\lambda >0$ whenever being  within a separation $\varrho>0$, called the reaction radius. 

\paragraph{Relation between rate constants for bimolecular reactions} \label{rate_relation}
   
Let $h$ denote the grid size of the compartment-based model (given by the standard RDME), and let $k_2/h^d$ (where $d$ is the dimension of physical space) denote the rate for two molecules of $A$ to react when located in the same compartment. The (standard) RDME may be seen as a formal approximation of the Doi model~\cite{isaacson2008relationship}, but it loses second-order reactions as $h\to 0$~ \cite{isaacson2006incorporating,isaacson2009reaction,erban2009stochastic}. In~\cite{isaacson2013convergent} a \textit{convergent} RDME has been developed, which allows second-order reactions of molecules located in different compartments and thereby does not lose bimolecular reactions as $h\to 0$. The \textit{standard} RDME may be interpreted as an asymptotic approximation of the \textit{convergent} RDME for $\varrho/h \ll 1$~\cite{isaacson2013convergent}, where $\varrho$ is the reaction radius of the Doi model. In this case of a comparatively large compartment size, particles can be expected to only react when being located in the same box, and an adequate choice of the binding rate constant $k_2$ is given by $k_2=\lambda |B_{\varrho}|$, where $|B_{\varrho}|$ denotes the volume of the $d$-dimensional sphere of radius $\varrho$. For our one-dimensional domain $[0,L]$ (i.e, $d=1$) we thus choose $k_2= 2 \, \varrho \, \lambda$ for the reaction-rate constant of the compartment-based model with sufficiently large compartment size $h$. Note that, on the other hand, $h$ has to be small, $h\ll L$, to ensure an appropriate level of spatial resolution~\cite{erban2020stochastic}.

\subsection{Model comparison} \label{sec:comparison_dim}

\begin{figure}
\centering
\includegraphics[width=0.7\textwidth]{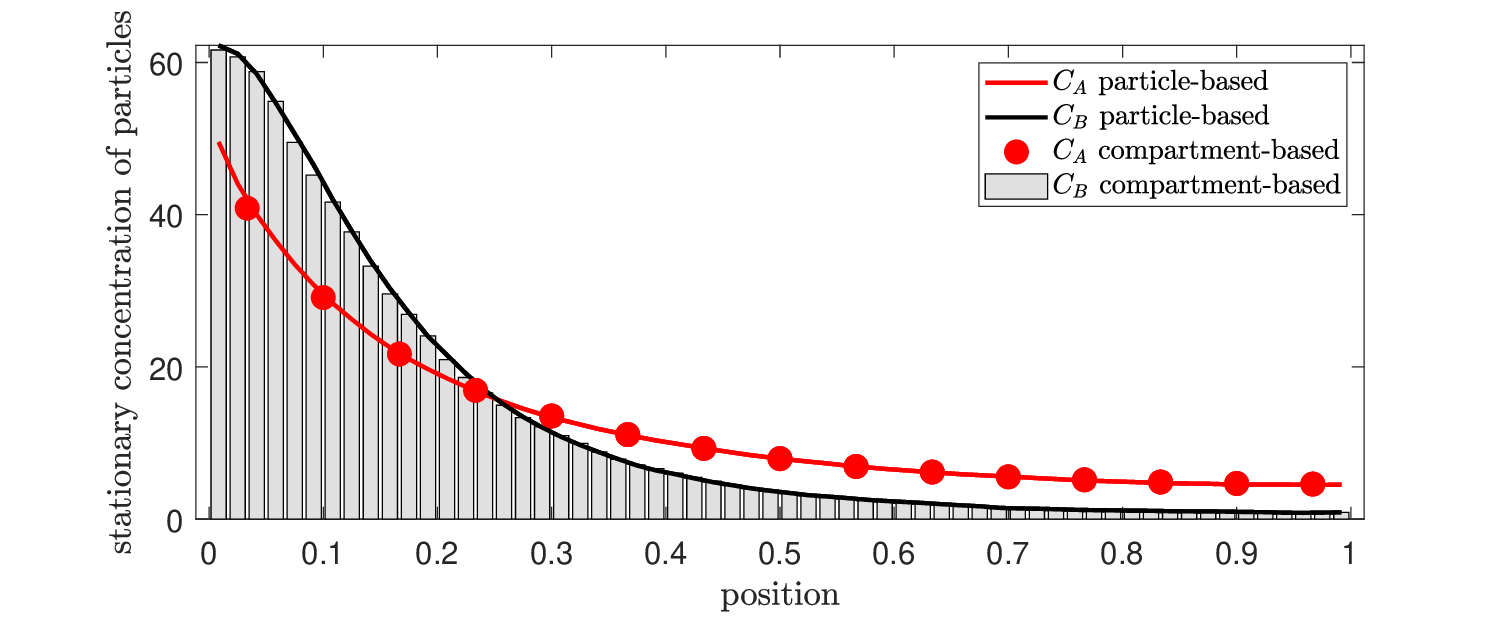}
\caption{{\textbf{Steady-state solution for the morphogen system with the dimerization reaction.}} {\it Comparison between Brownian dynamics $($solid lines$)$ and generalized {\rm mgRDME} model $($dots/ bars$)$ for dimerization system~\eqref{reactions_dimerization}. The parameters are given in \eqref{parameter_values_dim}, together with $L=1$, $K_A=15$ and $K_B=60$.}
}
\label{dimerization}
\end{figure}

In contrast to the first-order reaction-diffusion system of Section~\ref{sec2}, analytical insights for the dimerization system are relatively limited, so our studies solely rely on computational simulations. Figure~\ref{dimerization} shows the steady-state gradients of molecules of $A$ and $B$, estimated from long-term stochastic simulations (using the Gillespie stochastic simulation algorithm for the generalized mgRDME and temporal discretization for the Brownian dynamics model). The parameters are chosen as
\begin{equation}\label{parameter_values_dim}
k_1=50, \;\, \lambda= 5, \;\, \varrho=0.02, \;\, 
k_2=2 \, \lambda \, \varrho = 0.2, \;\, k_3=2, \;\, D_A=0.16, \;\, D_B=0.01. \;
\end{equation}
We observe a close agreement of the generalized compartment-based model with the particle-based dynamics for $K_A=15$ and $K_B=60$. In contrast, for the standard RDME the agreement is worse, as we will see in the next section.  

\subsubsection{Multi-objective optimization}

For the spatial error in the distribution of molecules of $B$, we define the function $\err(K_B,\gamma)$ in analogy to~\eqref{err1}, that is, we choose the $L^1$-distance between the steady-state solutions of the compartment-based model and of the particle-based model. Let $\bar{A}_{\text{total}}$ and $\bar{B}_{\text{total}}$ be the total number of molecules of $A$ and $B$, respectively, of the compartment-based dimerization system at steady state. In contrast to the first-order reaction system of Section~\ref{sec2}, we lack analytical expression for these quantities and can only estimate them via long-term numerical simulation.  The cost function is defined similarly to~\eqref{cost}, using the total propensity for diffusion and reaction events at steady state:
\begin{equation} \label{cost_dim}
c(K_B,\gamma) 
:= 2\,\frac{D_A}{h_A^2}\bar{A}_{\text{total}} +
2\,\frac{D_B}{h_B^2}\bar{B}_{\text{total}} + k_1 +k_2\,\bar{A}_{\text{total}} (\bar{A}_{\text{total}} -1) + k_3\bar{B}_{\text{total}}\,, 
\end{equation}
which is quadratic in the total number of molecules of $A$ at steady state. 
\begin{figure}
\centering
\includegraphics[width=1\textwidth]{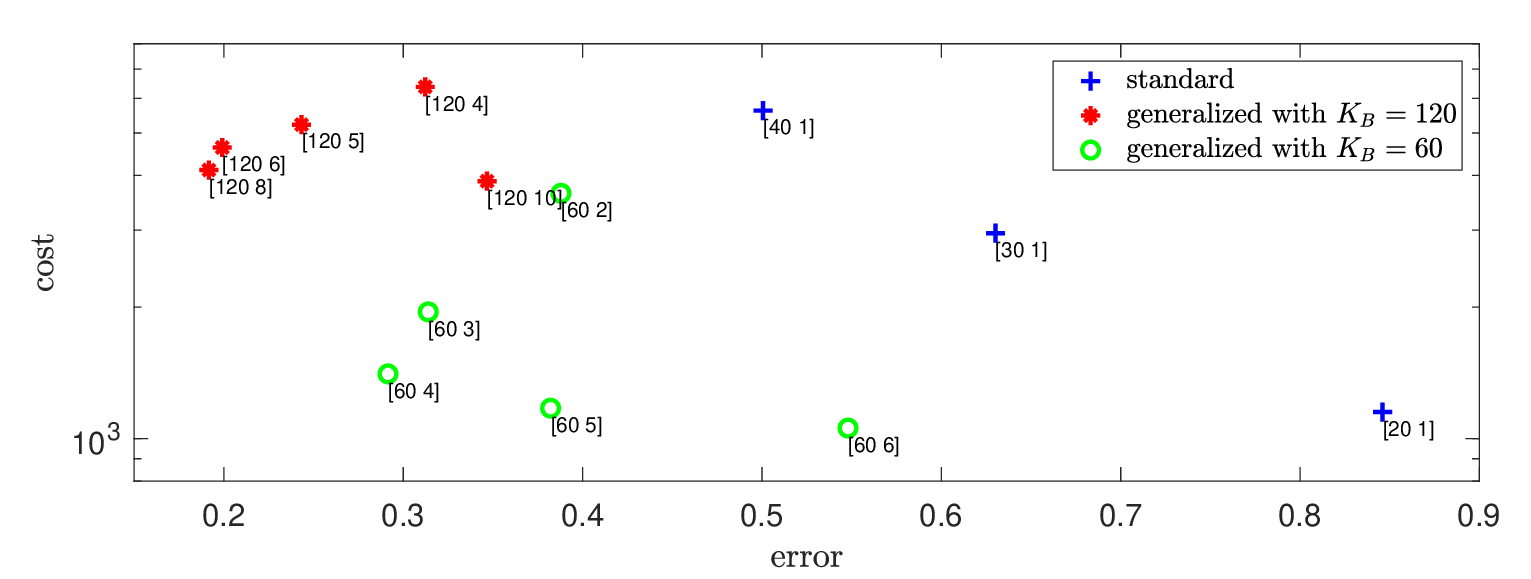}
\caption{{\textbf{Error and cost for dimerization.}} {\it Cost $c(K_B,\gamma)$ as defined in~\eqref{cost_dim}  and error $\err(K_B,\gamma)$ given by~\eqref{err1} for different values of $K_B$ and $\gamma=K_B/K_A$ $($given by labels of the form $[K_B \; \gamma]\,)$. The parameters are given in \eqref{parameter_values_dim} and $L=1$. The results are calculated as averages over long-term simulations of length $T=5000$.}}
    \label{fig:MOO_dim}
\end{figure}
Figure~\ref{fig:MOO_dim} illustrates the error and cost values across different combinations of $K_B$ and $\gamma$. We see that the error is \textit{non-monotone} in $\gamma$, which differs from the behavior seen in Figure~\ref{fig:MOO} for the first-order reaction-diffusion system. It seems that there is an \textit{optimal} $K_A \approx 15 =120/8 =60/4$, where the error in the spatial distribution of molecules of $B$ is minimal. Next, we study the relationship between the parameter $K_A$ and the error in the $A$-population.

\subsubsection{Error analysis for species $A$}

\begin{figure}
\centering
\begin{subfigure}[b]{0.49\textwidth}
\includegraphics[width=1\textwidth]{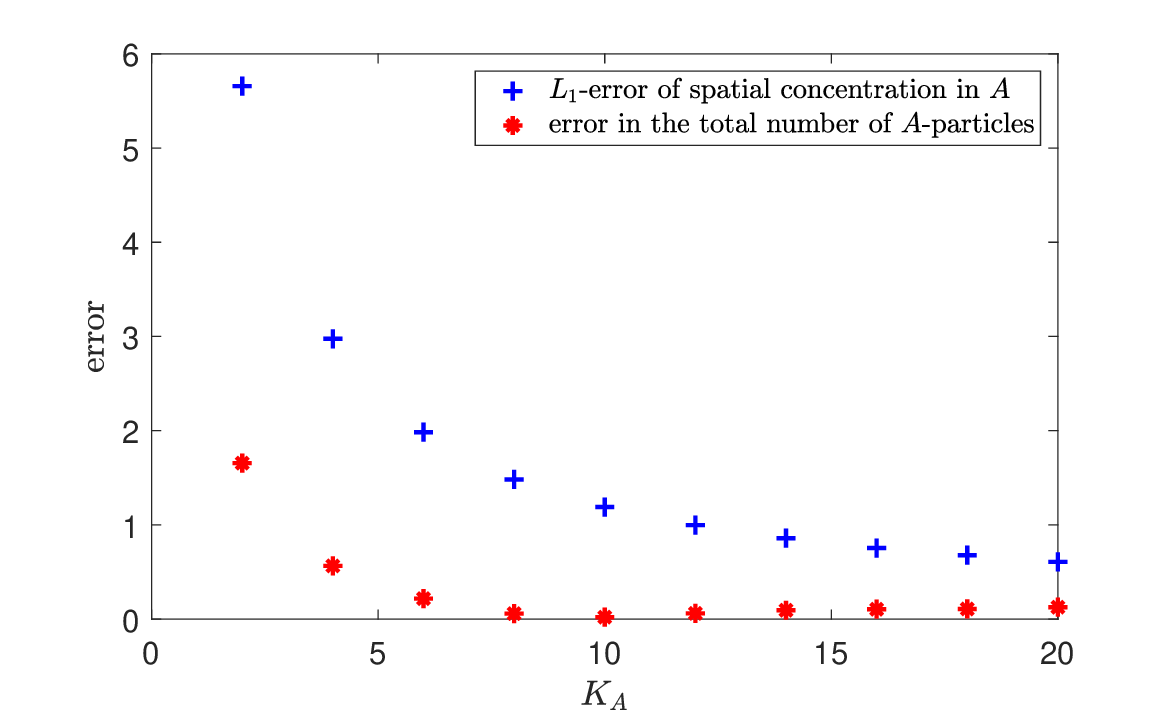}
\caption{Error in $A$ for $K_A\in\{2,4,...,20\}$}
\label{fig:errorA}
\end{subfigure}
\begin{subfigure}[b]{0.49\textwidth}
\includegraphics[width=1\textwidth]{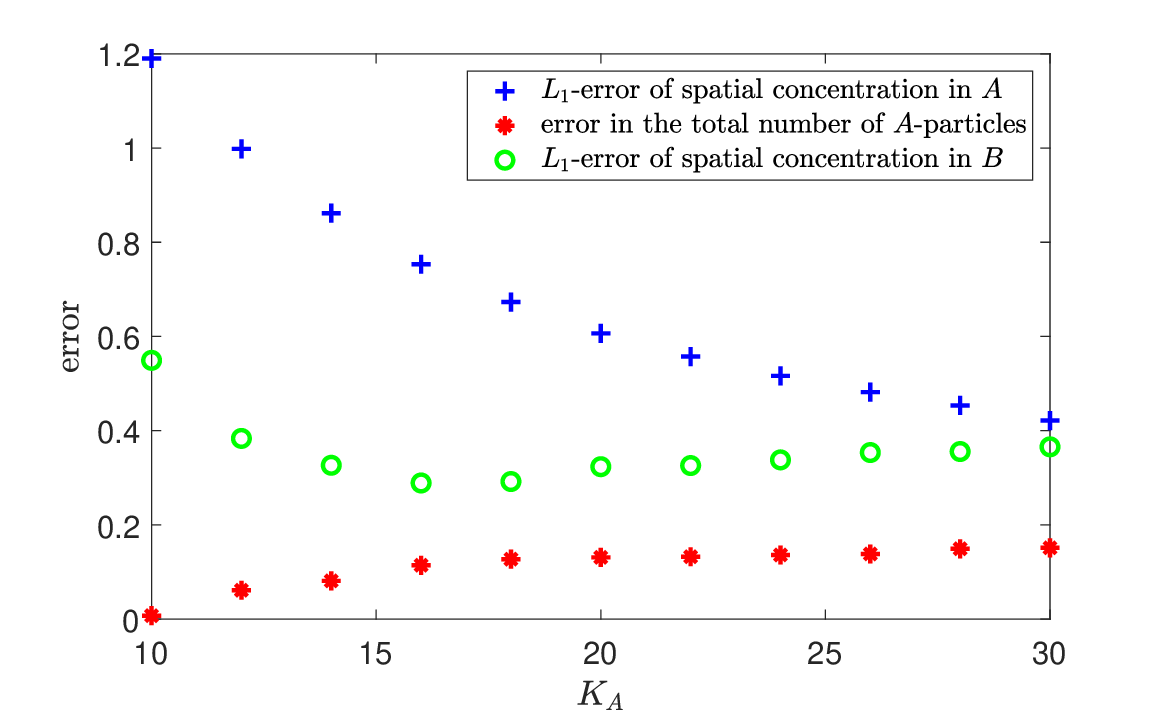}
\caption{Error in $A$ and $B$ for $K_A\in\{10,12,...,20\}$}
\label{fig:errorAB}
\end{subfigure}
\caption{{ \textbf{Error in dependence on $K_A$.}} {\rm (a)} {\it Spatial error $\err_A(K_A)$ given by~\eqref{err1_A} $($blue$)$ and difference $\text{\normalfont  err}_{\text{total}}(K_A)$ in the total number of molecules of $A$ defined in \eqref{e_total} $($red$)$ for different values of $K_A$. {\rm (b)} zoom-in,  additionally contains the spatial error in~$B$, $\err(K_B,\gamma)$ defined in~\eqref{err1}, for fixed $K_B=60$ and $\gamma=K_B/K_A$. Parameters given in \eqref{parameter_values_dim}, $L=1$. Long-term simulation of length $T=10^5$.}}
\label{fig:errorA_dim}
\end{figure}

Given the steady-state solution $\bar{\mathbf A}_{h_A}=(\bar{A}_1,\bar{A}_2,\dots,\bar{A}_{K_A})$ of the generalized mgRDME model for the dimerization system, we define 
\begin{equation}
\bar{a}_{h_A}(x) := \bar{A}_i/h_A \quad  \mbox{ for } x\in [(i-1)h_A,ih_A), \quad i=1,2,\dots,K_A,
\end{equation} 
as an approximation of the position-dependent steady-state concentration. Let further $\bar{a}(x)$ be the steady-state solution of the particle-based Brownian-dynamics dimerization model (these functions can be approximately determined via numerical simulation).
We define the spatial error in $A$ in analogy to~\eqref{err1}:
\begin{equation}\label{err1_A}
\err_A(K_A) := ||\bar{a} -\bar{a}_h||_{L^1} = \int_0^L |\bar{a}(x)-\bar{a}_h(x)| \, \mbox{d}x.
\end{equation}
Moreover, we consider the difference in the total number of molecules of $A$ at steady state:
\begin{equation}\label{e_total}
\err_{\text{total}}(K_A) := \left| \bar{A}_{\text{total}} - \int_0^L \bar{a}(x) \, \mbox{d}x\right|. 
\end{equation}
As opposed to the first-order system, this quantity can here be non-zero. Figure~\ref{fig:errorA_dim} shows a non-monotone relationship between the number $K_A$ of compartments and the error $\err_{\text{total}}(K_A)$ in the total number of molecules of $A$. The non-monotony is due to a too small grid size $h_A$ for large $K_A$ which contrasts with the condition $\varrho/h \ll 1$ necessary for a good approximation of second-order reactions, see also the paragraph on the relation between the rates on page \pageref{rate_relation}. The spatial error $\err_A(K_A) $ in $A$, however, decreases monotonically with increasing $K_A$ because a higher level of spatial resolution dominates the effect of losing second-order reactions for the parameter values under consideration. The error in $B$ (also depicted in Figure~\ref{fig:errorAB}) results from a combination of the two errors in $A$ and minimizes for $K_A\approx 16$, which is in consistency with the observations from Figure~\ref{fig:MOO_dim}. The standard compartment-based model would mean to choose $K_A=K_B=60$, inducing a comparatively large error both in the steady-state distribution of molecules of $B$ and in the total number of molecules of $A$.  

We conclude that the generalized mgRDME model clearly outperforms the standard RDME model for the fast-slow morphogen gradient system with dimerization. While in the setting of first-order reactions, it is reasonable to choose a specific ratio $\gamma=\sqrt{D_A/D_B}$ between the spatial compartment numbers, the non-linear case of dimerization suggests a fixed compartment size $h_A$ for molecules of $A$, regardless of the spatial resolution for $B$.  Instead, the reaction radius $\varrho$ and the rate constant $\lambda$ of the particle-based Brownian dynamics model, in conjunction with the diffusion constant $D_A$, dictate the optimal value for $h_A$.

\vskip 1mm

\begin{remark}[Second-order moments for dimerization]
Unlike the first-order reaction-diffusion model discussed in Section~$\ref{sec2}$, the variance of the particle numbers does here not align with the mean. Consequently, one could also take the variance as a value for comparing the models and defining an error. A numerical comparison of mean and variance for the different model types in application to the dimerization system is given in Appendix~$\ref{app_variance}$.  
\end{remark}

\section{Discussion}\label{sec:conclusion}

We examined the mgRDME as a generalized compartment-based stochastic model for reaction-diffusion kinetics, allowing adapted compartment sizes for different chemical species involved in the system~\cite{cao2014stochastic}. When employed in the context of morphogen gradient formation, the mgRDME appears to be very useful because the signaling molecules $A$ typically diffuse at a significantly higher rate than the morphogen molecules $B$, and this diffusion rate is a deciding factor for the choice of compartment sizes. In comparison to the `ground truth' particle-based Brownian dynamics simulations, the mgRDME clearly outperformed the standard RDME in terms of accuracy and reduction in numerical cost, both for the first-order reaction network~\eqref{reactions_linear} and for the considered dimerization system~\eqref{reactions_dimerization}.

Interestingly, however, the structural dependence of error and cost on the compartment sizes is different for the two settings. In Section~\ref{sec2}, the optimal ratio between grid sizes hinges on the relationship between the diffusion rates. Conversely, in the system involving second-order dimerization in Section~\ref{sec3}, an optimal grid size for species $A$ can be identified (in dependence on its diffusion constant) regardless of the grid size allocated to $B$. In particular, the approximation error may be reduced for the dimerization system by choosing a smaller compartment size for $B$ while maintaining the fixed optimal compartment size for $A$ undergoing dimerization. This goes in line with the observations for standard RDMEs, where very small grid sizes result in the elimination of the second-order reactions in three-dimensional simulations. The mgRDME circumvents this issue by allowing larger compartment sizes for species involved in second-order reactions.
At the same time, the increase in numerical cost caused by the higher spatial resolution is limited, since it is selectively applied solely to the slow species. 

Our work shows that the mgRDME is a valuable framework for studying reaction-diffusion systems with multiple scales in diffusion speed of involved molecules. The multi-grid network enables the mgRDME to more effectively accommodate the inherent conditions of the system under consideration, surpassing the adaptability of the standard RDME. The models with different diffusion constants naturally take into account variations in the sizes of simulated biomolecules, which can range from small calcium ions to relatively large vesicles
in neurotransmission dynamics~\cite{ernst2022variance,ernst2023rate},
or from relatively small $G$-actin monomers to larger $F$-actin filaments in simulations of actin dynamics~\cite{Zhuravlev:2010:DAT,Dobramysl:2016:SEI}. Other applications of mgRDME include pattern formation based on Turing instability which requires different diffusion constants~\cite{Turing:1952:CBM}.

In this work, we have solely considered dynamics in one-dimensional domains -- a restriction which is well justified in the context of morphogen gradient formation. Nevertheless, it would be interesting to extend our studies to reaction-diffusion systems in higher-dimensional domains, as well as to other reaction networks. While our analysis has been based on comparison of steady-state morphogen gradients, another way to compare the RDME and mgRDME could be by comparing the first collision and mean reaction times in these models~\cite{Li:2018:RTT}, which can be achieved by analyzing the random walks on lattices~\cite{Montroll:1969:RWL}. Also a comparison of time-dependent solutions (in contrast to steady-state distributions investigated here) is another interesting topic for future research on mgRDME models of morphogen gradients~\cite{saunders2009pays}.

\appendix
\section{Appendix}
\label{sec:Appendix}

\subsection{Time-dependent solution of the Brownian dynamics model}
\label{sectimeBD}

Let us consider that initially there are no molecules of $A$ and $B$ in the system, {\it i.e.}
\begin{equation}
a(x,0) \equiv b(x,0) \equiv 0.
\label{initcondition}    
\end{equation}
To solve the PDE~(\ref{pde1}) with initial condition~(\ref{initcondition}) and boundary conditions~(\ref{boundcond}), we apply the Fourier transform
$$
\widehat{a}(\xi,t)
=
\int_{-\infty}^{\infty}
a(x,t)
\, e^{-i \xi x}
\,
\mbox{d}
x.
$$
We get
$$
\frac{\partial \widehat{a}}{\partial t} (\xi,t)
=
- (D_A \, \xi^2 + k_2) \, \widehat{a}(\xi,t)
+ 2 k_1. 
$$
Solving this ODE with the initial condition $\widehat{a}(\xi,0)=0$, we obtain
\begin{equation}
\widehat{a}(\xi,t)
=
\frac{2k_1}{D_A \, \xi^2 +k_2}
-
\frac{2k_1 \, \exp\!\big[ - \left(D_A \, \xi^2 +k_2\right) t \, \big]}{D_A \, \xi^2 +k_2}.
\label{fourtransforma}    
\end{equation}
Since the Fourier transform of the convolution is the product of Fourier transforms, we can calculate the Fourier inverse of~(\ref{fourtransforma}) as follows
$$
a(x,t)
=
\bar{a}(x)
-
\frac{\exp[-k_2 t]}{2 \sqrt{D_A t \pi}}
\int_{-\infty}^{\infty}
\bar{a}(y)
\,
\exp\!\left[ - \frac{(x-y)^2}{4 D_A t} \right]
\,
\mbox{d}
y,
$$
where $\bar{a}(x)$ is given by~(\ref{assteadystatesol}). In total we get
\begin{align}
a(x,t)
= & 
\frac{k_1}{\sqrt{D_A \, k_2}} \,
\exp \!\left[ 
-\sqrt{\frac{k_2}{D_A}} \, |x|
\right]  \nonumber \\
& 
- \frac{\exp[-k_2 t]}{2 \sqrt{D_A t \pi}}
\int_{-\infty}^{\infty}
\frac{k_1}{\sqrt{D_A \, k_2}} \,
\exp \!\left[ 
-\sqrt{\frac{k_2}{D_A}} \, |y|
\right]
\,
\exp\!\left[ - \frac{(x-y)^2}{4 D_A t} \right]
\,
\mbox{d}
y
\label{solaxt}
\end{align}
for the average number of $A$-molecules in the interval $[x,x+\mbox{d}x]$ at time $t$.

\subsection{'Ground truth' solution: The effect of the boundary} \label{secboundBD}

Here, we derive the `ground truth' solution of Section~\ref{sec:ground_truth} for zero-flux boundary conditions. To do that, we will solve the ODEs~\eqref{ststeq1}--\eqref{ststeq2} on the interval $[-L,L]$ with zero-flux boundary conditions
\begin{equation}
\frac{\mbox{d} \bar{a}}{\mbox{d} x}
(-L)
=
\frac{\mbox{d} \bar{a}}{\mbox{d} x}
(L)
=
\frac{\mbox{d} \bar{b}}{\mbox{d} x}
(-L)
=
\frac{\mbox{d} \bar{b}}{\mbox{d} x}
(L)
=
0.
\label{boundcondODE2}
\end{equation}
Then the solution to the steady-state equation~\eqref{ststeq1} is 
$$\bar{a}(x) 
= 
\frac{k_1}{\sqrt{D_A \, k_2}} \,
\exp \!\left[ 
-\sqrt{\frac{k_2}{D_A}} \, |x|
\right]
\frac{
1
+
\exp \!\left[ 
2 \sqrt{\frac{k_2}{D_A}} \, \Big(|x|-L\Big)
\right]
}{1-\exp[-2 L \sqrt{k_2/D_A}]},
$$
which for $L\to\infty$ agrees with \eqref{assteadystatesol}, independently of the parameter values.

\subsection{Second-order moments for dimerization} \label{app_variance}

\begin{figure}
    \centering
	\begin{subfigure}[b]{0.49\textwidth}
    \includegraphics[width=0.9\textwidth]{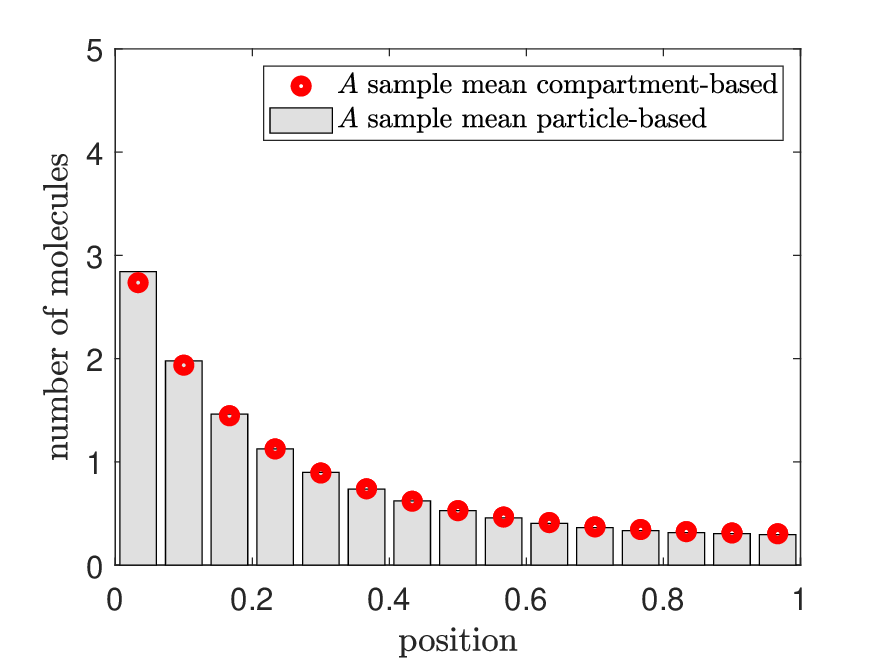}
    \caption{mean $A$}
    \end{subfigure}
    \hfill
    	\begin{subfigure}[b]{0.49\textwidth}
    \includegraphics[width=0.9\textwidth]{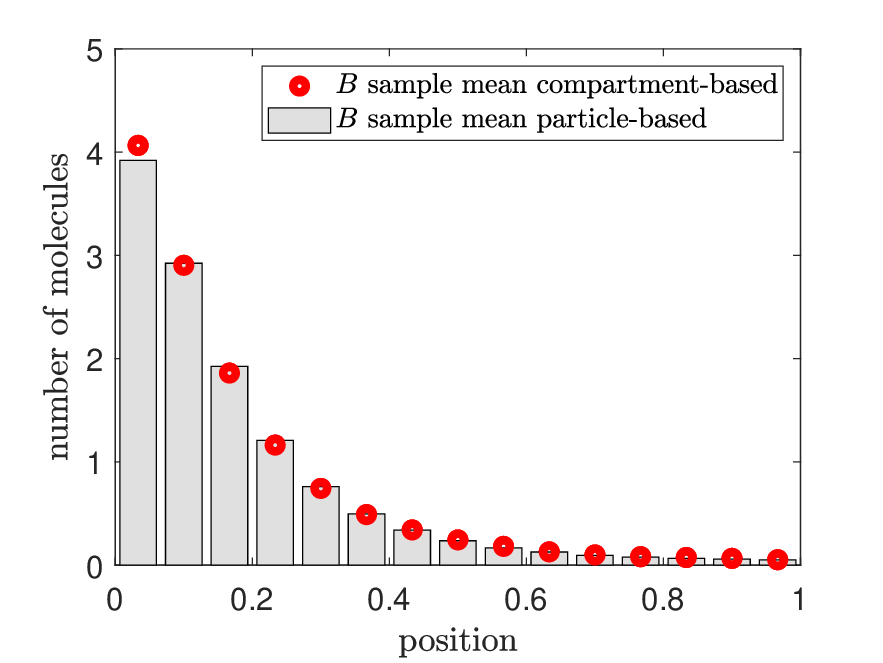}
    \caption{mean $B$}
    \end{subfigure}
    	\begin{subfigure}[b]{0.49\textwidth}
    \includegraphics[width=0.9\textwidth]{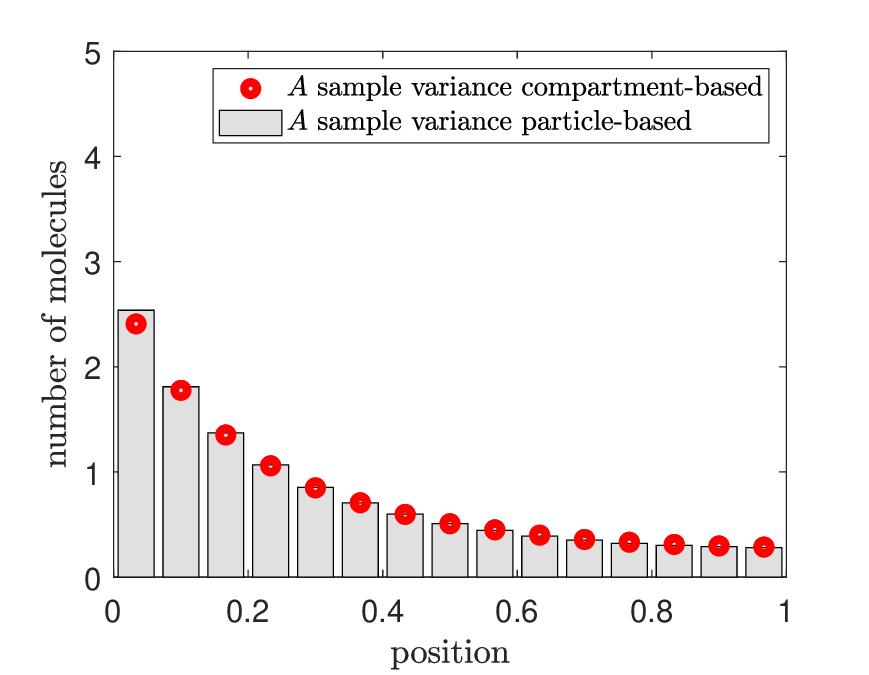}
    \caption{variance $A$}
    \end{subfigure}
    \hfill
    	\begin{subfigure}[b]{0.49\textwidth}
    \includegraphics[width=0.9\textwidth]{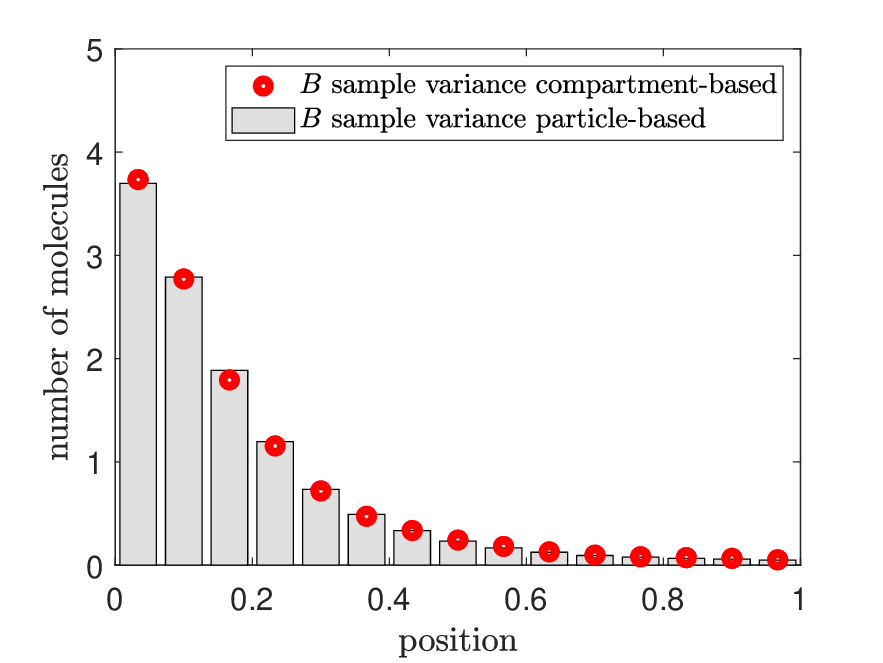}
    \caption{variance $B$}
    \end{subfigure}
    \caption{{ \textbf{Comparison of mean and variance for dimerization.}} {\it Mean/variance of the number of molecules of \normalfont(a)/(c) \it species $A$ and \normalfont (b)/(d) \it species $B$ at steady state estimated by long-term simulation $(T=2000)$ of the compartment-based model $($red circles$)$ and of the particle-based Brownian dynamics model $($grey bars$)$. Rate constants given in~\eqref{parameter_values_dim}, $L=1$, $K_A=K_B=15$ $(\gamma =1$, $h=1/15)$.} 
    }
    \label{fig:dimerization_mean_variance}
\end{figure}

Means and variances of the numbers of molecules of $A$ and $B$ for the dimerization system are plotted in Figure~\ref{fig:dimerization_mean_variance}, sampled from long-term simulations of the (standard) compartment-based model and of the particle-based Brownian dynamics model, choosing boxes of size $h=1/15$. We observe a good agreement of the two model types. 

\vskip 4mm

\noindent
{\bf Funding.}
This work was supported by the Engineering and Physical Sciences Research Council, grant
number EP/V047469/1, awarded to Radek Erban. We further acknowledge the support of Deutsche Forschungsgemeinschaft through CRC~1114 and through Germany’s Excellence Strategy--The Berlin Mathematics Research Center~MATH+ (EXC-2046/1, project 390685689), awarded to Stefanie Winkelmann. 

\bibsep=0.1em


\begin{thebibliography}{58}
\ifx \bisbn   \undefined \def \bisbn  #1{ISBN #1}\fi
\ifx \binits  \undefined \def \binits#1{#1}\fi
\ifx \bauthor  \undefined \def \bauthor#1{#1}\fi
\ifx \batitle  \undefined \def \batitle#1{#1}\fi
\ifx \bjtitle  \undefined \def \bjtitle#1{#1}\fi
\ifx \bvolume  \undefined \def \bvolume#1{\textbf{#1}}\fi
\ifx \byear  \undefined \def \byear#1{#1}\fi
\ifx \bissue  \undefined \def \bissue#1{#1}\fi
\ifx \bfpage  \undefined \def \bfpage#1{#1}\fi
\ifx \blpage  \undefined \def \blpage #1{#1}\fi
\ifx \burl  \undefined \def \burl#1{\textsf{#1}}\fi
\ifx \doiurl  \undefined \def \doiurl#1{\url{https://doi.org/#1}}\fi
\ifx \betal  \undefined \def \betal{\textit{et al.}}\fi
\ifx \binstitute  \undefined \def \binstitute#1{#1}\fi
\ifx \binstitutionaled  \undefined \def \binstitutionaled#1{#1}\fi
\ifx \bctitle  \undefined \def \bctitle#1{#1}\fi
\ifx \beditor  \undefined \def \beditor#1{#1}\fi
\ifx \bpublisher  \undefined \def \bpublisher#1{#1}\fi
\ifx \bbtitle  \undefined \def \bbtitle#1{#1}\fi
\ifx \bedition  \undefined \def \bedition#1{#1}\fi
\ifx \bseriesno  \undefined \def \bseriesno#1{#1}\fi
\ifx \blocation  \undefined \def \blocation#1{#1}\fi
\ifx \bsertitle  \undefined \def \bsertitle#1{#1}\fi
\ifx \bsnm \undefined \def \bsnm#1{#1}\fi
\ifx \bsuffix \undefined \def \bsuffix#1{#1}\fi
\ifx \bparticle \undefined \def \bparticle#1{#1}\fi
\ifx \barticle \undefined \def \barticle#1{#1}\fi
\bibcommenthead
\ifx \bconfdate \undefined \def \bconfdate #1{#1}\fi
\ifx \botherref \undefined \def \botherref #1{#1}\fi
\ifx \url \undefined \def \url#1{\textsf{#1}}\fi
\ifx \bchapter \undefined \def \bchapter#1{#1}\fi
\ifx \bbook \undefined \def \bbook#1{#1}\fi
\ifx \bcomment \undefined \def \bcomment#1{#1}\fi
\ifx \oauthor \undefined \def \oauthor#1{#1}\fi
\ifx \citeauthoryear \undefined \def \citeauthoryear#1{#1}\fi
\ifx \endbibitem  \undefined \def \endbibitem {}\fi
\ifx \bconflocation  \undefined \def \bconflocation#1{#1}\fi
\ifx \arxivurl  \undefined \def \arxivurl#1{\textsf{#1}}\fi
\csname PreBibitemsHook\endcsname

\bibitem[\protect\citeauthoryear{Fange and Elf}{2006}]{Fange:2006:NMP}
\begin{barticle}
\bauthor{\bsnm{Fange}, \binits{D.}},
\bauthor{\bsnm{Elf}, \binits{J.}}:
\batitle{Noise-induced {M}in phenotypes in {E}. coli}.
\bjtitle{PLoS Computational Biology}
\bvolume{2}(\bissue{6}),
\bfpage{637}--\blpage{648}
(\byear{2006})
\end{barticle}
\endbibitem

\bibitem[\protect\citeauthoryear{Earnest et~al.}{2015}]{Earnest:2015:TWM}
\begin{barticle}
\bauthor{\bsnm{Earnest}, \binits{T.M.}},
\bauthor{\bsnm{Lai}, \binits{J.}},
\bauthor{\bsnm{Chen}, \binits{K.}},
\bauthor{\bsnm{Hallock}, \binits{M.J.}},
\bauthor{\bsnm{Williamson}, \binits{J.R.}},
\bauthor{\bsnm{Luthey-Schulten}, \binits{Z.}}:
\batitle{Toward a whole-cell model of ribosome biogenesis: kinetic modeling of
  {SSU} assembly}.
\bjtitle{Biophysical journal}
\bvolume{109}(\bissue{6}),
\bfpage{1117}--\blpage{1135}
(\byear{2015})
\end{barticle}
\endbibitem

\bibitem[\protect\citeauthoryear{Denizot et~al.}{2019}]{Denizot:2019:SCS}
\begin{barticle}
\bauthor{\bsnm{Denizot}, \binits{A.}},
\bauthor{\bsnm{Arizono}, \binits{M.}},
\bauthor{\bsnm{N\"agerl}, \binits{U.}},
\bauthor{\bsnm{Soula}, \binits{H.}},
\bauthor{\bsnm{Berry}, \binits{H.}}:
\batitle{Simulation of calcium signaling in fine astrocytic processes: Effect
  of spatial properties on spontaneous activity}.
\bjtitle{PLoS Computational Biology}
\bvolume{15}(\bissue{8}),
\bfpage{1006795}
(\byear{2019})
\end{barticle}
\endbibitem

\bibitem[\protect\citeauthoryear{Dobramysl et~al.}{2016}]{Dobramysl:2016:PMM}
\begin{barticle}
\bauthor{\bsnm{Dobramysl}, \binits{U.}},
\bauthor{\bsnm{R\"udiger}, \binits{S.}},
\bauthor{\bsnm{Erban}, \binits{R.}}:
\batitle{Particle-based multiscale modeling of calcium puff dynamics}.
\bjtitle{Multiscale Modelling and Simulation}
\bvolume{14}(\bissue{3}),
\bfpage{997}--\blpage{1016}
(\byear{2016})
\end{barticle}
\endbibitem

\bibitem[\protect\citeauthoryear{Winkelmann and
  Sch{\"u}tte}{2016}]{winkelmann2016spatiotemporal}
\begin{botherref}
\oauthor{\bsnm{Winkelmann}, \binits{S.}},
\oauthor{\bsnm{Sch{\"u}tte}, \binits{C.}}:
The spatiotemporal master equation: Approximation of reaction--diffusion
  dynamics via {M}arkov state modeling.
Journal of Chemical Physics
\textbf{145}(21)
(2016)
\end{botherref}
\endbibitem

\bibitem[\protect\citeauthoryear{Zhuravlev et~al.}{2010}]{Zhuravlev:2010:DAT}
\begin{barticle}
\bauthor{\bsnm{Zhuravlev}, \binits{P.}},
\bauthor{\bsnm{Der}, \binits{B.}},
\bauthor{\bsnm{Papoian}, \binits{G.A.}}:
\batitle{Design of active transport must be highly intricate: A possible role
  of myosin and ena/{VASP} for {G}-actin transport in filopodia}.
\bjtitle{Biophysical Journal}
\bvolume{98},
\bfpage{1439}--\blpage{1448}
(\byear{2010})
\end{barticle}
\endbibitem

\bibitem[\protect\citeauthoryear{Erban et~al.}{2014}]{Erban:2014:MSR}
\begin{barticle}
\bauthor{\bsnm{Erban}, \binits{R.}},
\bauthor{\bsnm{Flegg}, \binits{M.}},
\bauthor{\bsnm{Papoian}, \binits{G.}}:
\batitle{Multiscale stochastic reaction-diffusion modelling: application to
  actin dynamics in filopodia}.
\bjtitle{Bulletin of Mathematical Biology}
\bvolume{76}(\bissue{4}),
\bfpage{799}--\blpage{818}
(\byear{2014})
\end{barticle}
\endbibitem

\bibitem[\protect\citeauthoryear{Winkelmann
  et~al.}{2021}]{winkelmann2021mathematical}
\begin{barticle}
\bauthor{\bsnm{Winkelmann}, \binits{S.}},
\bauthor{\bsnm{Zonker}, \binits{J.}},
\bauthor{\bsnm{Sch{\"u}tte}, \binits{C.}},
\bauthor{\bsnm{Conrad}, \binits{N.D.}}:
\batitle{Mathematical modeling of spatio-temporal population dynamics and
  application to epidemic spreading}.
\bjtitle{Mathematical Biosciences}
\bvolume{336},
\bfpage{108619}
(\byear{2021})
\end{barticle}
\endbibitem

\bibitem[\protect\citeauthoryear{Erban and Chapman}{2020}]{erban2020stochastic}
\begin{bbook}
\bauthor{\bsnm{Erban}, \binits{R.}},
\bauthor{\bsnm{Chapman}, \binits{S.J.}}:
\bbtitle{Stochastic Modelling of Reaction--{D}iffusion Processes}.
\bpublisher{Cambridge University Press},
\blocation{Cambridge}
(\byear{2020})
\end{bbook}
\endbibitem

\bibitem[\protect\citeauthoryear{Winkelmann and
  Sch{\"u}tte}{2020}]{winkelmann2020stochastic}
\begin{bbook}
\bauthor{\bsnm{Winkelmann}, \binits{S.}},
\bauthor{\bsnm{Sch{\"u}tte}, \binits{C.}}:
\bbtitle{Stochastic Dynamics in Computational Biology}
vol. \bseriesno{645}.
\bpublisher{Springer},
\blocation{Cham, Switzerland}
(\byear{2020})
\end{bbook}
\endbibitem

\bibitem[\protect\citeauthoryear{Flegg et~al.}{2012}]{Flegg:2012:TRM}
\begin{barticle}
\bauthor{\bsnm{Flegg}, \binits{M.}},
\bauthor{\bsnm{Chapman}, \binits{J.}},
\bauthor{\bsnm{Erban}, \binits{R.}}:
\batitle{The two-regime method for optimizing stochastic reaction-diffusion
  simulations}.
\bjtitle{Journal of the Royal Society Interface}
\bvolume{9}(\bissue{70}),
\bfpage{859}--\blpage{868}
(\byear{2012})
\end{barticle}
\endbibitem

\bibitem[\protect\citeauthoryear{Montefusco et~al.}{2023}]{montefusco2023route}
\begin{barticle}
\bauthor{\bsnm{Montefusco}, \binits{A.}},
\bauthor{\bsnm{Sch{\"u}tte}, \binits{C.}},
\bauthor{\bsnm{Winkelmann}, \binits{S.}}:
\batitle{A route to the hydrodynamic limit of a reaction--diffusion master
  equation using gradient structures}.
\bjtitle{SIAM Journal on Applied Mathematics}
\bvolume{83}(\bissue{2}),
\bfpage{837}--\blpage{861}
(\byear{2023})
\end{barticle}
\endbibitem

\bibitem[\protect\citeauthoryear{Flegg et~al.}{2015}]{Flegg:2015:CMC}
\begin{barticle}
\bauthor{\bsnm{Flegg}, \binits{M.}},
\bauthor{\bsnm{Hellander}, \binits{S.}},
\bauthor{\bsnm{Erban}, \binits{R.}}:
\batitle{Convergence of methods for coupling of microscopic and mesoscopic
  reaction-diffusion simulations}.
\bjtitle{Journal of Computational Physics}
\bvolume{289},
\bfpage{1}--\blpage{17}
(\byear{2015})
\end{barticle}
\endbibitem

\bibitem[\protect\citeauthoryear{Kang and Erban}{2019}]{Kang:2019:MSR}
\begin{barticle}
\bauthor{\bsnm{Kang}, \binits{H.}},
\bauthor{\bsnm{Erban}, \binits{R.}}:
\batitle{Multiscale stochastic reaction-diffusion algorithms combining {M}arkov
  chain models with stochastic partial differential equations}.
\bjtitle{Bulletin of Mathematical Biology}
\bvolume{81}(\bissue{8}),
\bfpage{3185}--\blpage{3213}
(\byear{2019})
\end{barticle}
\endbibitem

\bibitem[\protect\citeauthoryear{Isaacson}{2009}]{isaacson2009reaction}
\begin{barticle}
\bauthor{\bsnm{Isaacson}, \binits{S.A.}}:
\batitle{The reaction--diffusion master equation as an asymptotic approximation
  of diffusion to a small target}.
\bjtitle{SIAM Journal on Applied Mathematics}
\bvolume{70}(\bissue{1}),
\bfpage{77}--\blpage{111}
(\byear{2009})
\end{barticle}
\endbibitem

\bibitem[\protect\citeauthoryear{Erban and Chapman}{2009}]{erban2009stochastic}
\begin{barticle}
\bauthor{\bsnm{Erban}, \binits{R.}},
\bauthor{\bsnm{Chapman}, \binits{S.J.}}:
\batitle{Stochastic modelling of reaction--diffusion processes: algorithms for
  bimolecular reactions}.
\bjtitle{Physical Biology}
\bvolume{6}(\bissue{4}),
\bfpage{046001}
(\byear{2009})
\end{barticle}
\endbibitem

\bibitem[\protect\citeauthoryear{Hellander
  et~al.}{2012}]{hellander2012reaction}
\begin{barticle}
\bauthor{\bsnm{Hellander}, \binits{S.}},
\bauthor{\bsnm{Hellander}, \binits{A.}},
\bauthor{\bsnm{Petzold}, \binits{L.}}:
\batitle{Reaction--diffusion master equation in the microscopic limit}.
\bjtitle{Physical Review E}
\bvolume{85}(\bissue{4}),
\bfpage{042901}
(\byear{2012})
\end{barticle}
\endbibitem

\bibitem[\protect\citeauthoryear{Kang et~al.}{2012}]{kang2012new}
\begin{barticle}
\bauthor{\bsnm{Kang}, \binits{H.}},
\bauthor{\bsnm{Zheng}, \binits{L.}},
\bauthor{\bsnm{Othmer}, \binits{H.}}:
\batitle{A new method for choosing the computational cell in stochastic
  reaction--diffusion systems}.
\bjtitle{Journal of Mathematical Biology}
\bvolume{65},
\bfpage{1017}--\blpage{1099}
(\byear{2012})
\end{barticle}
\endbibitem

\bibitem[\protect\citeauthoryear{Isaacson}{2013}]{isaacson2013convergent}
\begin{botherref}
\oauthor{\bsnm{Isaacson}, \binits{S.A.}}:
A convergent reaction--diffusion master equation.
Journal of Chemical Physics
\textbf{139}(5)
(2013)
\end{botherref}
\endbibitem

\bibitem[\protect\citeauthoryear{Agbanusi and
  Isaacson}{2014}]{agbanusi2014comparison}
\begin{barticle}
\bauthor{\bsnm{Agbanusi}, \binits{I.C.}},
\bauthor{\bsnm{Isaacson}, \binits{S.A.}}:
\batitle{A comparison of bimolecular reaction models for stochastic
  reaction--diffusion systems}.
\bjtitle{Bulletin of Mathematical Biology}
\bvolume{76}(\bissue{4}),
\bfpage{922}--\blpage{946}
(\byear{2014})
\end{barticle}
\endbibitem

\bibitem[\protect\citeauthoryear{Hellander and
  Petzold}{2016}]{hellander2016reaction}
\begin{barticle}
\bauthor{\bsnm{Hellander}, \binits{S.}},
\bauthor{\bsnm{Petzold}, \binits{L.}}:
\batitle{Reaction rates for a generalized reaction--diffusion master equation}.
\bjtitle{Physical Review E}
\bvolume{93}(\bissue{1}),
\bfpage{013307}
(\byear{2016})
\end{barticle}
\endbibitem

\bibitem[\protect\citeauthoryear{Isaacson and
  Zhang}{2018}]{isaacson2018unstructured}
\begin{barticle}
\bauthor{\bsnm{Isaacson}, \binits{S.A.}},
\bauthor{\bsnm{Zhang}, \binits{Y.}}:
\batitle{An unstructured mesh convergent reaction--diffusion master equation
  for reversible reactions}.
\bjtitle{Journal of Computational Physics}
\bvolume{374},
\bfpage{954}--\blpage{983}
(\byear{2018})
\end{barticle}
\endbibitem

\bibitem[\protect\citeauthoryear{Doi}{1976a}]{doi1976stochastic}
\begin{barticle}
\bauthor{\bsnm{Doi}, \binits{M.}}:
\batitle{Stochastic theory of diffusion-controlled reaction}.
\bjtitle{Journal of Physics A: Mathematical and General}
\bvolume{9}(\bissue{9}),
\bfpage{1479}
(\byear{1976})
\end{barticle}
\endbibitem

\bibitem[\protect\citeauthoryear{Doi}{1976b}]{doi1976second}
\begin{barticle}
\bauthor{\bsnm{Doi}, \binits{M.}}:
\batitle{Second quantization representation for classical many-particle
  system}.
\bjtitle{Journal of Physics A: Mathematical and General}
\bvolume{9}(\bissue{9}),
\bfpage{1465}
(\byear{1976})
\end{barticle}
\endbibitem

\bibitem[\protect\citeauthoryear{Andrews et~al.}{2010}]{andrews2010detailed}
\begin{barticle}
\bauthor{\bsnm{Andrews}, \binits{S.S.}},
\bauthor{\bsnm{Addy}, \binits{N.J.}},
\bauthor{\bsnm{Brent}, \binits{R.}},
\bauthor{\bsnm{Arkin}, \binits{A.P.}}:
\batitle{Detailed simulations of cell biology with Smoldyn 2.1}.
\bjtitle{PLoS Computational Biology}
\bvolume{6}(\bissue{3}),
\bfpage{1000705}
(\byear{2010})
\end{barticle}
\endbibitem

\bibitem[\protect\citeauthoryear{Andrews}{2012}]{andrews2012spatial}
\begin{botherref}
\oauthor{\bsnm{Andrews}, \binits{S.S.}}:
Spatial and stochastic cellular modeling with the {S}moldyn simulator.
Bacterial Molecular Networks: Methods and Protocols,
519--542
(2012)
\end{botherref}
\endbibitem

\bibitem[\protect\citeauthoryear{Cao and Erban}{2014}]{cao2014stochastic}
\begin{barticle}
\bauthor{\bsnm{Cao}, \binits{Y.}},
\bauthor{\bsnm{Erban}, \binits{R.}}:
\batitle{Stochastic {T}uring patterns: analysis of compartment-based
  approaches}.
\bjtitle{Bulletin of Mathematical Biology}
\bvolume{76},
\bfpage{3051}--\blpage{3069}
(\byear{2014})
\end{barticle}
\endbibitem

\bibitem[\protect\citeauthoryear{Hellander and
  Hellander}{2020}]{Hellander:2020:HAR}
\begin{botherref}
\oauthor{\bsnm{Hellander}, \binits{S.}},
\oauthor{\bsnm{Hellander}, \binits{A.}}:
{Hierarchical algorithm for the reaction--diffusion master equation}.
Journal of Chemical Physics
\textbf{152}(3)
(2020)
\end{botherref}
\endbibitem

\bibitem[\protect\citeauthoryear{Li and Cao}{2012}]{li2012multiscale}
\begin{bchapter}
\bauthor{\bsnm{Li}, \binits{F.}},
\bauthor{\bsnm{Cao}, \binits{Y.}}:
\bctitle{Multiscale discretization for reaction diffusion systems}.
In: \bbtitle{Proceedings of the International Conference on Bioinformatics \&
  Computational Biology (BIOCOMP)},
p. \bfpage{242}
(\byear{2012})
\end{bchapter}
\endbibitem

\bibitem[\protect\citeauthoryear{Chatterjee et~al.}{2004}]{Chatterjee:2004:SAL}
\begin{barticle}
\bauthor{\bsnm{Chatterjee}, \binits{A.}},
\bauthor{\bsnm{Vlachos}, \binits{D.}},
\bauthor{\bsnm{Katsoulakis}, \binits{M.}}:
\batitle{Spatially adaptive lattice coarse-grained {M}onte {C}arlo simulations
  for diffusion of interacting molecules}.
\bjtitle{Journal of Chemical Physics}
\bvolume{121},
\bfpage{11420}--\blpage{11431}
(\byear{2004})
\end{barticle}
\endbibitem

\bibitem[\protect\citeauthoryear{Dai et~al.}{2008}]{Dai:2008:CGL}
\begin{barticle}
\bauthor{\bsnm{Dai}, \binits{J.}},
\bauthor{\bsnm{Seider}, \binits{W.}},
\bauthor{\bsnm{Sinno}, \binits{T.}}:
\batitle{Coarse-grained lattice kinetic {M}onte {C}arlo simulation of systems
  of strongly interacting particles}.
\bjtitle{Journal of Chemical Physics}
\bvolume{128},
\bfpage{194705}
(\byear{2008})
\end{barticle}
\endbibitem

\bibitem[\protect\citeauthoryear{Saunders and Howard}{2009}]{saunders2009pays}
\begin{barticle}
\bauthor{\bsnm{Saunders}, \binits{T.}},
\bauthor{\bsnm{Howard}, \binits{M.}}:
\batitle{When it pays to rush: interpreting morphogen gradients prior to
  steady-state}.
\bjtitle{Physical Biology}
\bvolume{6}(\bissue{4}),
\bfpage{046020}
(\byear{2009})
\end{barticle}
\endbibitem

\bibitem[\protect\citeauthoryear{Kicheva et~al.}{2007}]{kicheva2007kinetics}
\begin{barticle}
\bauthor{\bsnm{Kicheva}, \binits{A.}},
\bauthor{\bsnm{Pantazis}, \binits{P.}},
\bauthor{\bsnm{Bollenbach}, \binits{T.}},
\bauthor{\bsnm{Kalaidzidis}, \binits{Y.}},
\bauthor{\bsnm{Bittig}, \binits{T.}},
\bauthor{\bsnm{Julicher}, \binits{F.}},
\bauthor{\bsnm{Gonz{\'a}lez-Gait{\'a}n}, \binits{M.}}:
\batitle{Kinetics of morphogen gradient formation}.
\bjtitle{Science}
\bvolume{315}(\bissue{5811}),
\bfpage{521}--\blpage{525}
(\byear{2007})
\end{barticle}
\endbibitem

\bibitem[\protect\citeauthoryear{Bergmann et~al.}{2007}]{Bergmann:2007:PDB}
\begin{barticle}
\bauthor{\bsnm{Bergmann}, \binits{S.}},
\bauthor{\bsnm{Sandler}, \binits{O.}},
\bauthor{\bsnm{Sberro}, \binits{H.}},
\bauthor{\bsnm{Shnider}, \binits{S.}},
\bauthor{\bsnm{Schejter}, \binits{E.}},
\bauthor{\bsnm{Shilo}, \binits{B.}},
\bauthor{\bsnm{Barkai}, \binits{N.}}:
\batitle{Pre-steady-state decoding of the {B}icoid morphogen gradient}.
\bjtitle{PLoS Biology}
\bvolume{5}(\bissue{2}),
\bfpage{46}
(\byear{2007})
\end{barticle}
\endbibitem

\bibitem[\protect\citeauthoryear{Gregor et~al.}{2007}]{Gregor:2007:SND}
\begin{barticle}
\bauthor{\bsnm{Gregor}, \binits{T.}},
\bauthor{\bsnm{Wieschaus}, \binits{E.}},
\bauthor{\bsnm{McGregor}, \binits{A.}},
\bauthor{\bsnm{Bialek}, \binits{W.}},
\bauthor{\bsnm{Tank}, \binits{D.}}:
\batitle{Stability and nuclear dynamics of the bicoid morphogen gradient}.
\bjtitle{Cell}
\bvolume{130}(\bissue{1}),
\bfpage{141}--\blpage{152}
(\byear{2007})
\end{barticle}
\endbibitem

\bibitem[\protect\citeauthoryear{Turing}{1952}]{Turing:1952:CBM}
\begin{barticle}
\bauthor{\bsnm{Turing}, \binits{A.}}:
\batitle{The chemical basis of morphogenesis}.
\bjtitle{Philosophical Transactions of the Royal Society Lond.}
\bvolume{237},
\bfpage{37}--\blpage{72}
(\byear{1952})
\end{barticle}
\endbibitem

\bibitem[\protect\citeauthoryear{Murray}{2002}]{Murray:2002:MB2}
\begin{bbook}
\bauthor{\bsnm{Murray}, \binits{J.}}:
\bbtitle{{M}athematical {B}iology II: Spatial Models and Biomedical
  Applications}.
\bpublisher{Springer},
\blocation{New York}
(\byear{2002})
\end{bbook}
\endbibitem

\bibitem[\protect\citeauthoryear{Rogers and Schier}{2011}]{Rogers:2011:MGG}
\begin{barticle}
\bauthor{\bsnm{Rogers}, \binits{K.}},
\bauthor{\bsnm{Schier}, \binits{A.}}:
\batitle{Morphogen gradients: From generation to interpretation}.
\bjtitle{Annual Review of Cell and Developmental Biology}
\bvolume{27}(\bissue{1}),
\bfpage{377}--\blpage{407}
(\byear{2011})
\end{barticle}
\endbibitem

\bibitem[\protect\citeauthoryear{Shimmi and O'Connor}{2003}]{Shimmi:2003:PPT}
\begin{barticle}
\bauthor{\bsnm{Shimmi}, \binits{O.}},
\bauthor{\bsnm{O'Connor}, \binits{M.}}:
\batitle{Physical properties of {Tld, Sog, Tsg and Dpp} protein interactions
  are predicted to help create a sharp boundary in {B}mp signals during
  dorsoventral patterning of the {D}rosophila embryo}.
\bjtitle{Development}
\bvolume{130}(\bissue{19}),
\bfpage{4673}--\blpage{4682}
(\byear{2003})
\end{barticle}
\endbibitem

\bibitem[\protect\citeauthoryear{Shimmi et~al.}{2005}]{Shimmi:2005:FTD}
\begin{barticle}
\bauthor{\bsnm{Shimmi}, \binits{O.}},
\bauthor{\bsnm{Umulis}, \binits{D.}},
\bauthor{\bsnm{Othmer}, \binits{H.}},
\bauthor{\bsnm{Connor}, \binits{M.}}:
\batitle{Faciliated transport of a {D}pp/{S}cw heterodimer by {S}og/{T}sg leads
  to robust patterning of the {D}rosophila blastoderm embryo}.
\bjtitle{Cell}
\bvolume{120}(\bissue{6}),
\bfpage{873}--\blpage{886}
(\byear{2005})
\end{barticle}
\endbibitem

\bibitem[\protect\citeauthoryear{Wolpert et~al.}{2015}]{wolpert2015principles}
\begin{bbook}
\bauthor{\bsnm{Wolpert}, \binits{L.}},
\bauthor{\bsnm{Tickle}, \binits{C.}},
\bauthor{\bsnm{Arias}, \binits{A.M.}}:
\bbtitle{Principles of Development}.
\bpublisher{Oxford University Press, USA},
\blocation{Oxford}
(\byear{2015})
\end{bbook}
\endbibitem

\bibitem[\protect\citeauthoryear{Teimouri and
  Kolomeisky}{2018}]{teimouri2018discrete}
\begin{botherref}
\oauthor{\bsnm{Teimouri}, \binits{H.}},
\oauthor{\bsnm{Kolomeisky}, \binits{A.B.}}:
Discrete-state stochastic modeling of morphogen gradient formation.
Morphogen Gradients: Methods and Protocols,
199--221
(2018)
\end{botherref}
\endbibitem

\bibitem[\protect\citeauthoryear{Kolomeisky}{2011}]{kolomeisky2011formation}
\begin{barticle}
\bauthor{\bsnm{Kolomeisky}, \binits{A.B.}}:
\batitle{Formation of a morphogen gradient: Acceleration by degradation}.
\bjtitle{Journal of Physical Chemistry Letters}
\bvolume{2}(\bissue{13}),
\bfpage{1502}--\blpage{1505}
(\byear{2011})
\end{barticle}
\endbibitem

\bibitem[\protect\citeauthoryear{Kang and Redner}{1984}]{Kang:1984:SAK}
\begin{barticle}
\bauthor{\bsnm{Kang}, \binits{K.}},
\bauthor{\bsnm{Redner}, \binits{S.}}:
\batitle{Scaling approach for the kinetics of recombination processes}.
\bjtitle{Physical Review Letters}
\bvolume{52}(\bissue{12}),
\bfpage{955}
(\byear{1984})
\end{barticle}
\endbibitem

\bibitem[\protect\citeauthoryear{Montroll}{1969}]{Montroll:1969:RWL}
\begin{barticle}
\bauthor{\bsnm{Montroll}, \binits{E.}}:
\batitle{Random walks on lattices. {III}. {C}alculation of first-passage times
  with application to exciton trapping on photosynthetic units}.
\bjtitle{Journal of Mathematical Physics}
\bvolume{10}(\bissue{4}),
\bfpage{753}--\blpage{765}
(\byear{1969})
\end{barticle}
\endbibitem

\bibitem[\protect\citeauthoryear{Ben-Avraham and
  Redner}{1986}]{BenAvraham:1986:KNA}
\begin{barticle}
\bauthor{\bsnm{Ben-Avraham}, \binits{D.}},
\bauthor{\bsnm{Redner}, \binits{S.}}:
\batitle{Kinetics of $n$-species annihilation: Mean-field and
  diffusion-controlled limits}.
\bjtitle{Physical Review A}
\bvolume{34}(\bissue{1}),
\bfpage{501}
(\byear{1986})
\end{barticle}
\endbibitem

\bibitem[\protect\citeauthoryear{Jahnke and Huisinga}{2007}]{jahnke2007solving}
\begin{barticle}
\bauthor{\bsnm{Jahnke}, \binits{T.}},
\bauthor{\bsnm{Huisinga}, \binits{W.}}:
\batitle{Solving the chemical master equation for monomolecular reaction
  systems analytically}.
\bjtitle{Journal of Mathematical Biology}
\bvolume{54},
\bfpage{1}--\blpage{26}
(\byear{2007})
\end{barticle}
\endbibitem

\bibitem[\protect\citeauthoryear{Erban and Chapman}{2007}]{Erban:2007:RBC}
\begin{barticle}
\bauthor{\bsnm{Erban}, \binits{R.}},
\bauthor{\bsnm{Chapman}, \binits{S.J.}}:
\batitle{Reactive boundary conditions for stochastic simulations of
  reaction-diffusion processes}.
\bjtitle{Physical Biology}
\bvolume{4}(\bissue{1}),
\bfpage{16}--\blpage{28}
(\byear{2007})
\end{barticle}
\endbibitem

\bibitem[\protect\citeauthoryear{Engblom et~al.}{2009}]{engblom2009simulation}
\begin{barticle}
\bauthor{\bsnm{Engblom}, \binits{S.}},
\bauthor{\bsnm{Ferm}, \binits{L.}},
\bauthor{\bsnm{Hellander}, \binits{A.}},
\bauthor{\bsnm{L{\"o}tstedt}, \binits{P.}}:
\batitle{Simulation of stochastic reaction--diffusion processes on unstructured
  meshes}.
\bjtitle{SIAM Journal on Scientific Computing}
\bvolume{31}(\bissue{3}),
\bfpage{1774}--\blpage{1797}
(\byear{2009})
\end{barticle}
\endbibitem

\bibitem[\protect\citeauthoryear{Gillespie}{1977}]{gillespie1977exact}
\begin{barticle}
\bauthor{\bsnm{Gillespie}, \binits{D.T.}}:
\batitle{Exact stochastic simulation of coupled chemical reactions}.
\bjtitle{Journal of Physical Chemistry}
\bvolume{81}(\bissue{25}),
\bfpage{2340}--\blpage{2361}
(\byear{1977})
\end{barticle}
\endbibitem

\bibitem[\protect\citeauthoryear{Gillespie}{2007}]{gillespie2007stochastic}
\begin{barticle}
\bauthor{\bsnm{Gillespie}, \binits{D.T.}}:
\batitle{Stochastic simulation of chemical kinetics}.
\bjtitle{Annual Review of Physical Chemistry}
\bvolume{58},
\bfpage{35}--\blpage{55}
(\byear{2007})
\end{barticle}
\endbibitem

\bibitem[\protect\citeauthoryear{Lipkova et~al.}{2011}]{Lipkova:2011:ABD}
\begin{barticle}
\bauthor{\bsnm{Lipkova}, \binits{J.}},
\bauthor{\bsnm{Zygalakis}, \binits{K.}},
\bauthor{\bsnm{Chapman}, \binits{J.}},
\bauthor{\bsnm{Erban}, \binits{R.}}:
\batitle{Analysis of {B}rownian dynamics simulations of reversible bimolecular
  reactions}.
\bjtitle{SIAM Journal on Applied Mathematics}
\bvolume{71}(\bissue{3}),
\bfpage{714}--\blpage{730}
(\byear{2011})
\end{barticle}
\endbibitem

\bibitem[\protect\citeauthoryear{Isaacson}{2008}]{isaacson2008relationship}
\begin{barticle}
\bauthor{\bsnm{Isaacson}, \binits{S.A.}}:
\batitle{Relationship between the reaction--diffusion master equation and
  particle tracking models}.
\bjtitle{Journal of Physics A: Mathematical and Theoretical}
\bvolume{41}(\bissue{6}),
\bfpage{065003}
(\byear{2008})
\end{barticle}
\endbibitem

\bibitem[\protect\citeauthoryear{Isaacson and
  Peskin}{2006}]{isaacson2006incorporating}
\begin{barticle}
\bauthor{\bsnm{Isaacson}, \binits{S.A.}},
\bauthor{\bsnm{Peskin}, \binits{C.S.}}:
\batitle{Incorporating diffusion in complex geometries into stochastic chemical
  kinetics simulations}.
\bjtitle{SIAM Journal on Scientific Computing}
\bvolume{28}(\bissue{1}),
\bfpage{47}--\blpage{74}
(\byear{2006})
\end{barticle}
\endbibitem

\bibitem[\protect\citeauthoryear{Ernst et~al.}{2022}]{ernst2022variance}
\begin{barticle}
\bauthor{\bsnm{Ernst}, \binits{A.}},
\bauthor{\bsnm{Sch{\"u}tte}, \binits{C.}},
\bauthor{\bsnm{Sigrist}, \binits{S.J.}},
\bauthor{\bsnm{Winkelmann}, \binits{S.}}:
\batitle{Variance of filtered signals: Characterization for linear reaction
  networks and application to neurotransmission dynamics}.
\bjtitle{Mathematical Biosciences}
\bvolume{343},
\bfpage{108760}
(\byear{2022})
\end{barticle}
\endbibitem

\bibitem[\protect\citeauthoryear{Ernst et~al.}{2023}]{ernst2023rate}
\begin{barticle}
\bauthor{\bsnm{Ernst}, \binits{A.}},
\bauthor{\bsnm{Unger}, \binits{N.}},
\bauthor{\bsnm{Sch{\"u}tte}, \binits{C.}},
\bauthor{\bsnm{Walter}, \binits{A.M.}},
\bauthor{\bsnm{Winkelmann}, \binits{S.}}:
\batitle{Rate-limiting recovery processes in neurotransmission under sustained
  stimulation}.
\bjtitle{Mathematical Biosciences}
\bvolume{362},
\bfpage{109023}
(\byear{2023})
\end{barticle}
\endbibitem

\bibitem[\protect\citeauthoryear{Dobramysl et~al.}{2016}]{Dobramysl:2016:SEI}
\begin{barticle}
\bauthor{\bsnm{Dobramysl}, \binits{U.}},
\bauthor{\bsnm{Papoian}, \binits{G.}},
\bauthor{\bsnm{Erban}, \binits{E.}}:
\batitle{Steric effects induce geometric remodeling of actin bundles in
  filopodia}.
\bjtitle{Biophysical Journal}
\bvolume{110},
\bfpage{2066}--\blpage{2075}
(\byear{2016})
\end{barticle}
\endbibitem

\bibitem[\protect\citeauthoryear{Li et~al.}{2018}]{Li:2018:RTT}
\begin{barticle}
\bauthor{\bsnm{Li}, \binits{F.}},
\bauthor{\bsnm{Chen}, \binits{M.}},
\bauthor{\bsnm{Erban}, \binits{R.}},
\bauthor{\bsnm{Cao}, \binits{Y.}}:
\batitle{Reaction time for trimolecular reactions in compartment-based
  reaction-diffusion models}.
\bjtitle{Journal of Chemical Physics}
\bvolume{148}(\bissue{20}),
\bfpage{204108}
(\byear{2018})
\end{barticle}
\endbibitem

\end{thebibliography}
\end{document}